\documentclass[a4paper,10pt]{article}
\usepackage{lineno}
\pdfoutput=1 
\usepackage{authb   lk}
\usepackage{amssymb}
\usepackage{pdflscape}
\usepackage[T1]{fontenc} 
\usepackage{graphicx}
\usepackage{hyperref}
\usepackage{cancel}
\usepackage{amsmath}
\usepackage{cleveref}
\usepackage{subcaption}
\usepackage{graphicx}
\usepackage{multicol}
\usepackage{empheq}
\usepackage{multirow}
\usepackage[dvipsnames]{xcolor}

\def\arraystretch{1.5}

\def\e{\ensuremath{\epsilon}}
\def\d{\ensuremath{\partial}}

\def\e{\epsilon}

\renewcommand{\arraystretch}{1.5}

\def\be{\begin{equation}}
\def\ee{\end{equation}}
\def\bsp#1\esp{\begin{split}#1\end{split}}
\def\bpm{\begin{pmatrix}}
\def\epm{\end{pmatrix}}
\newcommand{\bea}{\begin{eqnarray}}
\newcommand{\eea}{\end{eqnarray}}
\def\lag{{\cal L}}

\def\gL{g_{\scriptscriptstyle L}}
\def\gR{g_{\scriptscriptstyle R}}
\def\gBL{g_{\scriptscriptstyle B-L}}
\def\gY{g_{\scriptscriptstyle Y}}

\def\vR{v_{\scriptscriptstyle R}}
\def\vL{v_{\scriptscriptstyle L}}
\def\tw{\theta_{\scriptscriptstyle W}}
\def\phw{\varphi_{\scriptscriptstyle W}}
\def\zw{\vartheta_{\scriptscriptstyle W}}

\usepackage[utf8x]{inputenc}

\usepackage{color}
\usepackage{hyperref}
\usepackage{fontenc}
\usepackage{pbox}
\usepackage{feynmp}

\usepackage{mathtools,nccmath}%
\usepackage{ etoolbox, xparse} 

\usepackage{array}
\newcolumntype{P}[1]{>{\centering\arraybackslash}p{#1}}

\graphicspath{{figs/},{figs/2l/},{figs/4l/},{figs/6l/}}

\def\arraystretch{1.5}

\def\e{\ensuremath{\epsilon}}
\def\d{\ensuremath{\partial}}

\def\e{\epsilon}

\def\be{\begin{equation}}
\def\ee{\end{equation}}
\def\bsp#1\esp{\begin{split}#1\end{split}}
\def\bpm{\begin{pmatrix}}
\def\epm{\end{pmatrix}}
\def\lag{{\cal L}}

\def\gL{g_{\scriptscriptstyle L}}
\def\gR{g_{\scriptscriptstyle R}}
\def\gBL{g_{\scriptscriptstyle B-L}}
\def\gY{g_{\scriptscriptstyle Y}}

\def\vR{v_{\scriptscriptstyle R}}
\def\vL{v_{\scriptscriptstyle L}}
\def\tw{\theta_{\scriptscriptstyle W}}
\def\phw{\varphi_{\scriptscriptstyle W}}
\def\zw{\vartheta_{\scriptscriptstyle W}}
\title{Exploring $0\nu\beta\beta$ and Leptogenesis in the Alternative Left-Right Model}

\author{Mariana Frank$^1$\footnote{mariana.frank@concordia.ca},  Chayan Majumdar$^2$\footnote{chayan@phy.iitb.ac.in},  P. Poulose$^{3,1}$\footnote{poulose@iitg.ac.in},  Supriya Senapati${^2}$\footnote{supriya@phy.iitb.ac.in}, and Urjit  A. Yajnik$^{2}$\footnote{yajnik@iitb.ac.in}}
\date{
{\small \it 
$^1$ Department of Physics,  
Concordia University, 7141 Sherbrooke St. West,
Montreal, Quebec, Canada H4B 1R6 \\
$^2$  Department of Physics, Indian Institute of Technology Bombay, Powai, Mumbai, Maharashtra 400 076, India\\
$^3$  Department of Physics, Indian Institute of Technology Guwahati, Assam 781 039, India
}\\[5mm]
\today
}

\begin{document}
	\maketitle
	\flushbottom
\begin{abstract}
We investigate the possibility of neutrinoless double beta decay ($0\nu\beta\beta$) and leptogenesis within the Alternative Left-Right  Model (ALRM). Unlike the usual left-right symmetric model, ALRM features a Majorana right-handed neutrino which does not carry any charge. Further, in this picture the down-type quark and the charged leptons receive mass through the additional left-handed  scalar field, rather than the usual doublet. Together, these features conspire to generate significant contributions to the $0\nu\beta\beta$ through vector-scalar ($WH$) mediation.  For moderate masses of the relevant charged Higgs boson ($M_{H_1^\pm}\sim 200$ GeV), the half-life of $T_{\frac{1}{2}}^{WH}$ is $\sim 3\times 10^{26}~{\rm yrs}$ for both the case of $^{76}$Ge and $^{136}$Xe, well within the sensitivity expected by
 future experiments.  Invoking the resonant leptogenesis, CP violation arising from the right-handed neutrino decay could  be the required order to generate the correct baryogenesis,  $\epsilon\sim 10^{-6}$, for small Dirac phases and without  any fine tuning.
\end{abstract}



\section{Introduction}
\label{Intro}

Neutrinos distinguished as an exclusively left chiral species in the electroweak interactions have provided an intriguing insight into the structure of fundamental interactions. The observation of neutrino oscillations, confirmed over several decades has signalled a key role for them, both in the flavour puzzle and in the understanding of conservation of baryon ($B$) and lepton ($L$) numbers, the global quantum numbers of the low energy Standard Model (SM). Study of such phenomena in turn can provide crucial hints towards a more complete and perhaps also a more elegant theory of elementary particles. Perhaps the most notable among these hints comes from the see-saw mechanism, linking the minuscule mass scale of the observed neutrinos to the grand unification scale, independently suggested by the running of the gauge coupling constants.  In all of the popular avatars of the see-saw mechanism, the appearance of newer right handed neutrino type species, and {\it inter alia}, the possibility of Majorana mass terms, is inevitable. The possibility of Majorana neutrinos 
gives rise to the neutrinoless double beta decay ($0\nu\beta\beta$) which has been pursued theoretically, and explored experimentally for many decades now. Further, this possibility invariably violates $L$, providing a natural way to address another intriguing issue in particle physics, the matter-anti-matter asymmetry of the Universe, by a mechanism known as leptogenesis. 

The observation of the proposed $0\nu\beta\beta$ will be a direct indication of Lepton Number Violation (LNV), as well as it can deliver  information regarding the absolute mass scale of the neutrino species.
Such LNV processes could arise due to the well explored mechanism with exchange of light Majorana neutrinos, but equally well from some new physics contribution, depending upon the beyond Standard Model (BSM) framework. Thus the study of this phenomenon provides crucial hints to constructing BSM. Among the various BSM frameworks, a minimal one incorporating right handed neutrinos and partially achieving the goals of unification is the class of left-right symmetric models (LRSM). 
  Extensive reviews on $0\nu\beta\beta$ decays in context of LRSM framework can be found in  \cite{Mohapatra:1980yp, Mohapatra:1981pm, Hirsch:1996qw, Tello:2010am, Chakrabortty:2012mh, Patra:2012ur, Awasthi:2013ff, Barry:2013xxa, Dev:2013vxa, Ge:2015yqa, 
Awasthi:2015ota, Halprin:1983ez, Albert:2014afa, Majumdar:2018eqz, Majumdar:2020owj}.
 
In the usual versions of the LRSM, the right-handed down-type quark teams up with the right-handed  up-type quark to form a doublet under $SU(2)_R$. Similarly, the right-handed electron pairs with the newly introduced right-handed neutrino to form a doublet under $SU(2)_R$. This scenario requires a triplet scalar field whose vacuum expectation value (VEV) provides the Majorana mass for the right-handed neutrino, as needed by the seesaw mechanism  \cite{Mohapatra:1977mj,Mohapatra:1974gc}. Since the simplest grand unified theory (GUT) based on $SU(5)$ has not been borne out by experiments, and in view of the massive nature of neutrinos, it has been natural to look to the left-right paradigm and any framework that can accommodate it in a fully unified model. All such grand unified theories naturally connect leptons and quarks. Grand unified theories originating from $E_6$ gauge group \cite{Hewett:1988xc,Langacker:1998tc,Ma:1986we},  embed a subgroup $SU(2)_L\times SU(2)_R\times U(1)$ with fermionic content capable of being identified with the proposed LRSM class of models.   However, an alternate charge assignment of the fermions in $E_6$ can lead to significant changes in the physical spectrum and the dynamics at low energies. We shall refer to this as Alternative Left-Right Model (ALRM)~\cite{Babu:1987kp,Ma:2010us}.

This model provides rich neutrino phenomenology. Several additional neutrino states are natural candidates for sterile neutrinos \cite{Frank:2004vg}.
In the usual LRSM, the important contribution to $0\nu\beta\beta$ decay comes from the $WW$ fusion channel production of the doubly charged scalar boson, which further decays to two electrons. In contrast, in the ALRM there are doublet scalar fields instead of the triplets. The peculiar partnership of the usual right-handed fermions (up type quarks and the charged leptons) with the exotic fermions leads to different constraints on the Yukawa couplings and VEVs. 
A consequence of these assignments is that some of the Yukawa couplings 
are significantly larger than in the case of LRSM,  generating potentially significant contributions to the $0\nu\beta\beta$ decay through charged Higgs exchange processes. Another important difference from the LRSM is that in the new set up, the right-handed gauge bosons do not couple the usual $u_R$ and $d_R$, meaning that $W_R$ does not contribute to $0\nu\beta\beta$ decay. In this paper, we  study such new physics contributions delivering sizable $0\nu\beta\beta$ estimation in context of ALRM framework.

Further, LNV is one of several broad paradigms for understanding the matter-anti-matter asymmetry of the Universe, generically dubbed baryon asymmetry of the Universe (BAU). The Majorana nature of the right-handed neutrino allows the possibility of leptogenesis. Again, the dynamics here is completely unlike the case of LRSM. First,  there is no influence of the SM Higgs boson, and the CP-asymmetry arises entirely through the decay of the right-handed neutrino to the left-handed charged scalar. Second, the couplings that influence this decay are different from those appearing in the usual $0\nu\beta\beta$ decay, and thus complementary to it. Third,  the possible washout aided by $W_R$ mediated decay and scattering processes present in the LRSM \cite{Frere:2008ct, Dev:2014iva} is completely absent in ALRM, owing to the absence of  interaction between $W_R$ and the right-handed neutrino. We therefore also study leptogenesis originating in this ALRM model which makes specific predictions about $0\nu\beta\beta$. 

We  explore these novel features originating in the alternate $E_6$ as possible explanations of two of the pressing puzzles of phenomenology. The article is organized as follows. In   \cref{sec:alrsm} we discuss the ALRM with its origins in $E_6$ and the existing constraints on the relevant couplings and masses.  For a more complete description of the model  one may refer to \cite{Frank:2019nid, Ashry:2013loa, Ashrythesis}.
In   \cref{sec:discussion} we move on to the calculations of $0\nu\beta\beta$ decay. There we present the possible channels and estimate their contribution identifying the important ones. In \cref{sec:leptogenesis} we discuss the leptogenesis, computing the CP-asymmetry arising from the decay of right-handed neutrinos in this model. We summarize our finding and present our conclusions in \cref{sec:conclusions}.

\section{The alternative left-right  model (ALRM)}
\label{sec:alrsm}
It is natural to seek embedding of the rank 5 group LRSM in Spin$(10)$ which happens minimally,  with an elegant assignment of fermions to the spinor $\mathbf{16}$. Later, it was realized in the context of superstring theory that Spin$(10)$ in turn needs to be embedded in $E_6$ (see \cite{Hewett:1988xc} for a review). However, it was further noted by Ma \cite{Ma:1986we} that an alternative embedding of low energy fermions in the representation  $\mathbf{27}$  is also possible. 
In this scenario, instead of partnering with the right-handed down-type quark, the right-handed up-quark joins with a new exotic colored fermion, $d^\prime_R$ (of the same charge as $d_R$), to form a doublet under the $SU(2)_{R^\prime}$. Similarly, the right-handed charged leptons partner with a new neutral fermion ($n_R$) to form a doublet under the same $SU(2)_{R^\prime}$. The right-handed down-type quark, $d_R$ and the right-handed neutrino, $\nu_R$, along with the left-handed degrees of freedom of the newly introduced fermions, $d_L^\prime$ and $n_L$, remain singlets under both $SU(2)_L$ and $SU(2)_{R^\prime}$. The model provides several additional neutrino states as natural candidates for sterile neutrinos \cite{Frank:2004vg}. Without supersymmetry, the model can provide  two scenarios for neutrino dark matter  \cite{Ma:2010us}. For this, an additional  $S$-symmetry is imposed, under which the lepton number is either $L=S-T_{3R}$,  in the Dark Left-Right Model (DLRM) \cite{Khalil:2009nb}, or $L=S+T_{3R}$, in the Dark Left-Right Model 2 (DLRM2) with a global $U(1)_S$ \cite{Khalil:2010yt}, duly extended to a local $U(1)_S$ in \cite{Kownacki:2017sqn}.
Finally, it was shown that the  partner of the right-handed electron, the scotino, is a viable DM candidate, consistent with all  constraints, within suitable parameter space region~\cite{Frank:2019nid}, and the model implications at colliders was explored. In this work, we shall work within the DLRM2, with global $S$ charge assignments as in   \cite{Ma:2012gb,  Ashry:2013loa, Ashrythesis}. 
With the right-handed neutrinos having $S=0$, there is a Majorana particle in the spectrum.

\begin{table}[h]
  \renewcommand{\arraystretch}{1.2}\setlength\tabcolsep{5pt}
  \begin{center}
  \small
    \begin{tabular}{ccc}
      Fields & {\small{$SU(3)_c\times SU(2)_L\times SU(2)_{R^\prime} \times U(1)_{B-L}$}}&$S$\\\hline\hline\\[-.4cm]
      $Q_L = \bpm u_L\\d_L\epm$ &
        $\big({\bf 3}, {\bf 2}, {\bf 1},  \frac16\big)$ & 0 \\[.4cm]
      $Q_R = \bpm u_R\\d_R'\epm$ &
        $\big({\bf 3}, {\bf 1}, {\bf 2},  \frac16\big)$ & $-\frac12$\\[.4cm]
      $d'_L$ & $\big({\bf 3}, {\bf 1}, {\bf 1}, -\frac13\big)$ & $-1$\\[.2cm]
      $d_R$  & $\big({\bf 3}, {\bf 1}, {\bf 1}, -\frac13\big)$ & 0\\[.2cm]
    \hline\\[-.4cm]
      $L_L = \bpm \nu_L\\e_L\epm$ &
        $\big({\bf 1}, {\bf 2}, {\bf 1},  -\frac12\big)$ & 0\\[.4cm]
      $L_R = \bpm n_R\\e_R\epm$  &
        $\big({\bf 1}, {\bf 1}, {\bf 2},  -\frac12\big)$ & $+\frac{1}{2}$\\[.4cm]
      $n_L$   & $\big({\bf 1}, {\bf 1}, {\bf 1}, 0\big)$ & +1\\[.2cm]
      $\nu_R$ & $\big({\bf 1}, {\bf 1}, {\bf 1}, 0\big)$ & 0\\ \hline
      $\phi = \bpm \phi_1^0&\phi_1^+\\ \phi_2^-&\phi^0_2 \epm$ &
         $\big({\bf 1}, {\bf 2}, {\bf 2}^*, 0\big)$ & $-\frac12$\\[.4cm]
      $\chi_L = \bpm\chi_L^+ \\\chi_L^0\epm$ &
         $\big({\bf 1}, {\bf 2}, {\bf 1}, \frac12\big)$ & 0 \\[.4cm]
      $\chi_R = \bpm\chi_R^+ \\\chi_R^0\epm$ &
         $\big({\bf 1}, {\bf 1}, {\bf 2}, \frac12\big)$ &$+\frac12$\\[.4cm]\hline
    \end{tabular}
    \caption{ALRM particle content, for one generation of fermions and for the Higgs fields 
 with  charge assignments under  $SU(3)_c\times SU(2)_L\times SU(2)_{R'} \times U(1)_{B-L}$
    (second column) and 
    the global $U(1)_S$ (third column). }
    \label{tab:content}
  \end{center}
\end{table}

The ALRM, arises from the breaking of $E_6$ starting with a substantially different embedding of the low energy fermion representations, and we shall refer to the symmetry of the resulting intermediate scale model as $SU(3)_c\times SU(2)_L\times SU(2)_{R^\prime} \times U(1)_{B-L}$, to distinguish it from the traditional LRSM. Along with this gauge group we consider DLRM2 type global $U(1)_S$ for the three generations of fermions, and several Higgs fields with the assigned charges as in table~\ref{tab:content}. Here $d^\prime$ and $n_{L,R}$ are exotic quarks and leptons, emerging from breaking of $E_6$.

With these assignments, the model Lagrangian includes, in addition to the standard gauge invariant kinetic terms
for all fields, a Yukawa interaction Lagrangian $\lag_{\rm Y}$, a Majorana mass term $\lag_{\rm M}$, and a scalar
potential $V_{\rm H}$, for both the bidoublet and the left and right doublet fields, $\chi_L$ and $\chi_R$. Accordingly, the Yukawa Lagrangian  and the Majorana mass terms in our model are given by
\be\label{eq:yuk}
-\lag_{\rm Y} = \bar Q_L { Y}^q \tilde\Phi Q_R
    + \bar Q_L {Y}^q_L \chi_L d_R
    + \bar Q_R { Y}^{q}_R\chi_R d'_L
    + \bar L_L { Y}^\ell \Phi L_R
    + \bar L_L { Y}^\ell_L \tilde\chi_L \nu_R
    + \bar L_R { Y}^\ell_R \tilde\chi_R n_L + {\rm h.c.} \ ,
\ee

\be
  -\lag_{\rm M}=m_N~\bar \nu_R^c\nu_R 
\ee
where  all the generation indices have been omitted for simplicity. The Yukawa
couplings ${ Y}$ are $3 \times 3$ matrices with generation labels as indices.
Likewise, the most general Higgs potential $V_{\rm H}$  is given by
\be\bsp
  V_{\rm H} = &
   -\mu_1^2 {\rm Tr} \big[\Phi^\dag \Phi\big]
   + \lambda_1 \big({\rm Tr}\big[\Phi^\dag \Phi\big]\big)^2
   + \lambda_2\ {\rm Tr}(\tilde \Phi^\dagger \Phi)\ {\rm Tr}(\Phi^\dag\tilde\Phi)
\\&
   -\mu_2^2 \big[\chi_L^\dag \chi_L + \chi_R^\dag \chi_R\big]+ \lambda_3 \Big[\big(\chi_L^\dag \chi_L\big)^2 +
         \big(\chi_R^\dag\chi_R\big)^2\Big]
   + 2 \lambda_4\ \big(\chi_L^\dag \chi_L\big)\ \big(\chi_R^\dag\chi_R\big) 
   \\&
   + 2 \alpha_1 {\rm Tr} \big[\Phi^\dag \Phi\big]
          \big[\chi_L^\dag \chi_L + \chi_R^\dag \chi_R\big]
            + 2 \alpha_2 \big[ \chi_L^\dagger \Phi\Phi^\dagger\chi_L+\chi_R^\dagger \Phi^\dagger\Phi\chi_R \big]
\\&
  + 2 \alpha_3 \big[ \chi_L^\dagger \tilde\Phi\tilde\Phi^\dagger\chi_L+\chi_R^\dagger \tilde\Phi^\dagger\tilde\Phi\chi_R \big]
   + \mu_3 \big[\chi_L^\dag \Phi \chi_R + \chi_R^\dag\Phi^\dag\chi_L\big] \ ,
\esp\label{eq:Hpot}
\ee
containing bilinear, trilinear  and quartic contributions. In the above expressions, the $SU(2)$ duals of the 
scalar fields are defined as
\be
\tilde\chi_{L(R)}=i\sigma^2 \chi_{L(R)}; \hspace{5mm} \tilde \Phi= \sigma^2\Phi\sigma^2 =  \bpm \phi_2^0*&-\phi_2^+\\-\phi_1^- & \phi_1^0* \epm 
\ee
where $\sigma^2$  is  the Pauli matrix in the standard notation.
The vacuum expectation value (VEV) acquired by the neutral component of $\chi_R$ breaks the 
$SU(2)_{R^\prime}\times U(1)_{B-L}$ symmetry down to the standard gauge symmetry, which is further broken to the electromagnetic gauge symmetry by the VEVs of the bidoublet and left-handed doublet fields.  The VEVs of these fields are given by
\be
  \langle \Phi  \rangle = \frac{1}{\sqrt{2}}\bpm 0&0\\0 & k \epm\ , \qquad
  \langle \chi_L\rangle = \frac{1}{\sqrt{2}}\bpm 0\\ \vL \epm\ , \qquad
  \langle \chi_R\rangle = \frac{1}{\sqrt{2}}\bpm 0\\ \vR \epm\ .
\label{eq:symbreak}
\ee
It is found that the VEV of the other neutral component of the bidoublet, $\phi^0_1$  remains zero \cite{Ashry:2013loa} which helps
to avoid the unwanted mixing between $d$ and $d'$ and also between $\nu_L$ and the scotino $n_R$.
This also decouples the $\phi^0_1$ field from the other neutral scalar fields, making its mass eigenstate  the same as the gauge eigenstate. 
With the complex neutral scalar fields expressed in terms of their  real
degrees of freedom, the fields are
\be\bsp
 \renewcommand{\arraystretch}{1.3}
  \phi_1^0 =&\ \frac{1}{\sqrt{2}}
     \Big[ \Re\{\phi^0_1\} + i\ \Im\{\phi^0_1\}\Big]\ ,
\\
  \phi_2^0 =&\ \frac{1}{\sqrt{2}}
     \Big[ k + \Re\{\phi^0_2\} + i\ \Im\{\phi^0_2\}\Big] \ ,
\\
  \chi_{L,R}^0 =&\ \frac{1}{\sqrt{2}}
     \Big[ v_{\scriptscriptstyle L,R} + \Re\{\chi^0_{L,R}\} +
     i\ \Im\{\chi^0_{L,R}\}\Big]\ .
\esp\ee
Then  the massive $CP$-even Higgs bosons  
$H_i^0$ (with $i=0,1,2,3$), the massive $CP$-odd Higgs bosons  $A_i^0$ (with
$i=1,2$) and the two massless Goldstone bosons $G_1^0$ and $G_2^0$ are expressed in terms of the gauge eigenstates, as
\be
\Im\{\phi_1^0\}= A_1^0,~~~~~~\Re\{\phi_1^0\}= H_1^0,
\label{eq:phi1}\ee

\be
 \bpm \Im\{\phi_2^0\}\\\Im\{\chi_L^0\}\\\Im\{\chi_R^0\}\epm =
 \bpm &&& \\ && U_{3\times 3}^{\rm A} & \\&&& \\ \epm
 \bpm A_2^0\\G_1^0 \\ G_2^0\epm
  \qquad\text{and} \qquad
 \bpm \Re\{\phi_2^0\}\\\Re\{\chi_L^0\}\\\Re\{\chi_R^0\}\epm =
 \bpm &&& \\ && U_{3\times 3}^{\rm H} & \\&&& \\ \epm
  \bpm H_0^0 \\ H_2^0\\H_3^0\epm   \ .
\label{eq:nh_mix}\ee
We refer  to Refs.~\cite{Frank:2019nid, Ashry:2013loa, Ashrythesis} for details and explicit expressions of the mixing matrices  
 $U_{3\times 3}^{\rm A}$ and $U_{3\times 3}^{\rm H}$, for the CP-odd and CP-even scalars. Pertinent for our further discussion are the charged scalar bosons and their interaction with the quarks and the leptons. The eight degrees of freedom in the charged  scalar sector of the unbroken symmetry, 
 $\phi_1^\pm$, $\phi_2^\pm$, $\chi_L^\pm$ and $\chi_R^\pm$ mix into two physical massive charged Higgs bosons $H_1^\pm$
and $H_2^\pm$, and  two massless Goldstone bosons $G_1^\pm$ and $G_2^\pm$ that are absorbed by the $W_L^\pm$ and $W_R^\pm$ gauge bosons,
\be\bsp
 \bpm \phi_1^\pm\\\chi_L^\pm\epm =
 \bpm \cos\beta & \sin\beta\\ -\sin\beta & \cos\beta\epm
 \bpm H_1^\pm\\G_1^\pm\epm \ , \ \
 \bpm \phi_2^\pm\\\chi_R^\pm\epm =
 \bpm \cos\zeta & \sin\zeta\\ -\sin\zeta & \cos\zeta\epm
 \bpm H_2^\pm\\G_2^\pm\epm \ ,
\esp\label{eq:ch_mix}\ee
with
\be
  \tan\beta = \frac{k}{\vL} \qquad\text{and}\qquad \tan\zeta = \frac{k}{\vR} \ .
\ee
Masses of the new physical charged bosons are given in terms of the parameters of the Lagrangian and the VEVs, as
\be
M_{H_1^\pm}=v^2~(\alpha_3-\alpha_2)-\frac{\mu_3}{\sqrt{2}}~\frac{v^2v_R}{kv_L},~~~~~~{\rm and},~~~~~~~
M_{H_2^\pm}=(k^2+v_R^2)~(\alpha_3-\alpha_2)-\frac{\mu_3}{\sqrt{2}}~\frac{(k^2+v_R^2)v_L}{kv_R} \, .
\ee
where $v^2=k^2+v_L^2$. The couplings of these charged scalars within  lepton and quark charged currents are relevant when discussing $0\nu\beta\beta$ and leptogenesis. 
To obtain these couplings, we examine the Yukawa couplings in more detail.
With the charged and neutral scalar degrees of freedom  as above, the gauge eigenstates of the bidoublet as well as of the two doublet scalar fields in the unitary gauge can be written in terms of the mass eigenstates as 
\be
\small
 \Phi=
 \bpm H_1^0+iA_1^0 & \cos\beta~H_1^+\\ \cos\zeta~H_2^- & k+\Re\{\phi_2^0\}+iU^A_{11}~A_2\epm,~~~~
 \chi_L=
 \bpm -\sin\beta~H_1^+\\
 v_L+\Re\{\chi_L^0\}+iU^A_{21}~A_2
 \epm,~~~
 \chi_R=
 \bpm -\sin\zeta~H_2^+\\
 v_R+\Re\{\chi_R^0\}+iU^A_{31}~A_2
 \epm .
 \label{unitary_fields}
\ee
With this, the leptonic part of the Yukawa Lagrangian can be expressed in terms of the mass eigenstates as
\bea
\scriptsize
\bar L_LY^\ell\Phi L_R& =&Y^\ell\left [
\big(H_1^0+iA_1^0\big)\,\bar \nu_Ln_R+\cos\beta~H_1^+\,\bar \nu_Le_R+ \cos\zeta~H_2^-\,\bar e_Ln_R\right. \nonumber\\
&&\left.+\big(k+\Re\{\phi_2^0\}+iU^A_{11}A_2\big)\,\bar e_Le_R +h.c. \right ] \nonumber \\[2mm]
\bar L_LY_L^\ell \tilde \chi_L\nu_R&=& Y_L^\ell\left [
\left(v_L+\Re\{ \chi^0_L\}-iU^A_{21}A_2\right)\,\bar \nu_L\nu_R+\sin\beta~H^-_1\,\bar e_L\nu_R \right ]\nonumber \\[2mm]
\bar L_RY^\ell_R\tilde \chi_Rn_L&=&Y_R^\ell\left [
\left(v_R+\Re\{ \chi^0_R\}-iU_{31}^A A_2\right)\,\bar n_R n_L+\sin\zeta~H_2^-\,\bar e_Rn_L\right ] \, .
\eea
And the Yukawa interactions connecting the  quarks are given by
\bea
\scriptsize
\bar Q_LY^q\tilde \Phi Q_R &=& Y^q\left [
\left(k+\Re\{\phi^0_2\}-iU^A_{11}A_2\right)\,\bar u_Lu_R-\cos\zeta\,H_2^+\,\bar u_Ld'_R\right.\nonumber \\
&&\left.{-\cos\beta~H_1^-\bar d_Lu_R}+\left(H_1^0-iA_1^0\right)\,\bar d_Ld'_R+h.c.\right ]\nonumber \\
\bar Q_LY^q_L\chi_Ld_R&=&Y^q_L\left [ -\sin\beta\,H_1^+\,\bar u_Ld_R+\left( v_L+\Re\{\chi_L^0\}+iU^A_{21}A_2\right)\,\bar d_Ld_R\right ] \nonumber \\
\bar Q_RY^q_R\chi_R d'_L&=&Y^q_R\left [-\sin\zeta\,H_2^+\,\bar u_Rd'_L+\left(v_R+\Re\{\chi^0_R\}+iU_{31}^AA_2 \right)\,d'_L\right ] \, .
\eea
Notice that the same Yukawa interactions are responsible for the quark and lepton masses, including those of the light neutrinos, but up and down quarks, and leptons and neutrinos, respectively, acquire masses from different Yukawa couplings:
\be
m_u=\frac{1}{\sqrt{2}}Y^q k, ~~~~ m_d=\frac{1}{\sqrt{2}}Y^q_Lv_L,~~~~m_\ell=\frac{1}{\sqrt{2}}Y^\ell k,~~~m_\nu = \frac{1}{m_N}\left(\frac{1}{\sqrt{2}}Y_L^\ell v_L\right)^2\, ,
\label{eq:quarkmass}
\ee
where $u$ and $d$ refer to the up-type quarks $u,~c,~t$ and the down-type quarks $d,~s,~b$, respectively; and $\ell=e,~\mu,~\tau$ are the charged leptons. While in the case of usual LRSM one has the liberty to choose $v_L$ (the VEV of the left-handed triplet scalar field there) as small as one likes, including setting it to zero, here the down type quarks acquire  mass proportional $v_L$. This, along with the maximum allowed value of the bottom-Yukawa coupling, while preserving perturbativity, leads to the choice
\be
Y^{b}_L\sim 1; \hspace{10mm}v_L \sim 5 ~\mathrm{GeV}
\ee
so that from $k^2+v_L^2=246^2~{\rm GeV}^2$, we still have $k\sim 246~{\rm GeV}$.  
With these values of the VEVs  the remaining Yukawa couplings relevant to our discussion are given by
\be
Y^u = \frac{\sqrt{2} m_u}{k}\sim 1.26 \times 10^{-5},~~~~~Y^d_L= \frac{\sqrt{2} m_d}{v_L}\sim 1.33 \times 10^{-3},
\ee
for the first generation quarks, and 
\be
Y^e = \frac{\sqrt{2} m_e}{k}\sim 2.93 \times 10^{-6},~~~~Y^\mu= \frac{\sqrt{2} m_\mu}{k}\sim 6.08 \times 10^{-4},
~~~~Y^\tau= \frac{\sqrt{2} m_\tau}{k}\sim 1.02 \times 10^{-2}.
\label{eq:quarkcoupling}
\ee
for the charged leptons. 
Further, with the light neutrino mass $m_\nu\sim 0.01~{\rm eV}$ and the heavy neutrino mass $m_N\sim 10$ TeV, the corresponding Yukawa couplings  are 
\be
Y_L^\ell =\frac{ \sqrt{2m_\nu m_N}}{v_L}\sim 8.96 \times 10^{-5} \, .
\label{eq:leptoncoupling}
\ee
Similarly,  the ratio of the VEVs, $\tan\beta = \frac{k}{v_L}\sim \frac{246}{5}\approx 49.2$. Or,
\be
\cos\beta \sim 0.0203~~~~~~~~~{\rm and}~~~~~~~~~~~\sin\beta\sim 1.
\label{eq:cosine}
\ee
The gauge sector of the model does not really make any impact on our discussion, as the $W_R$ does not couple to $\nu_R$ or $d_R$,   but  connects the right-handed charged leptons and up-type quarks to $d'_R$ and $n_R$, respectively,  the exotic fermions. However, for completeness we  briefly  discuss gauge boson masses below. As usual, the spontaneous breaking of the left-right symmetry generates the boson masses  and possibly induces their mixing. While the charged gauge bosons $W_L$ and $W_R$ could mix in the usual LRSM case, in the ALRM  they do not mix as $\langle\phi_1^0
\rangle = 0$. In the present case, their masses are given by
\be
  M_{W_L}    = \frac12 \gL \sqrt{k^2+\vL^2} \equiv \frac12 \gL v
 \qquad\text{and}\qquad
  M_{W_R} = \frac12 \gR \sqrt{k^2+\vR^2}~.
\label{eq:mw_mwp}\ee
In the neutral sector, the gauge boson mass-squared  matrix is written, in the
$(B_\mu, W_{L\mu}^3, W_{R\mu}^3)$ basis, as
\be
  ({\cal M}^0_V)^2 = \frac14 \bpm
    \gBL^2\ (\vL^2+\vR^2)  & -\gBL\ \gL\ \vL^2    & -\gBL\ \gR\ \vR^2\\
   -\gBL\ \gL\ \vL^2       &  \gL^2\ v^2          & -\gL\ \gR\ k^2\\
   -\gBL\ \gR\ \vR^2       & -\gL\ \gR\ k^2       &  \gR^2\ (k^2+\vR^2)
  \epm \ .
\ee
It can be diagonalized through three rotations that mix the $B$, $W_L^3$ and
$W_R^3$ bosons into the massless photon $A$ and massive $Z$ and $Z'$ states,
\renewcommand{\arraystretch}{1.}
\be
  \bpm B_\mu\\ W_{L\mu}^3\\ W_{R\mu}^3\epm = 
  \bpm \cos {\phw} & 0 & -\sin {\phw}\\ 0 & 1 & 0\\ \sin {\phw} & 0 & \cos {\phw} \epm
  \bpm \cos {\tw} & -\sin {\tw} & 0\\ \sin {\tw} & \cos {\tw} & 0\\ 0 & 0 & 1 \epm
  \bpm 1 & 0 & 0\\ 0 & \cos {\zw} & -\sin {\zw}\\ 0 & s_{\zw} & \cos {\zw} \epm
  \bpm A_\mu\\ Z_\mu\\ Z^\prime_\mu\epm  \ ,
\ee
The $\phw$-rotation mixes the $B$ and $W_R^3$ bosons into the hypercharge boson
$B'$ and its orthogonal combination as generated by the breaking of $SU(2)_{R^\prime}\times U(1)_{B-L}$ into to the
hypercharge group $U(1)_Y$. The usual electroweak mixing between $B'$ and $W^3_L$  generating the photon, $A$ and the orthogonal combination is achieved by the $\tw$-rotation, and finally the $\zw$-rotation is related to $Z-Z^\prime$ mixing, which is strongly constrained. The mixing angles $\phw$ and  $\tw$ are related to the gauge couplings in a straightforward manner, 
\be\bsp
 & \sin {\phw} = \frac{\gBL}{\sqrt{\gBL^2+\gR^2}} = \frac{\gY}{\gR}
   \qquad\text{and}\qquad
   \sin {\tw}  = \frac{\gY}{\sqrt{\gL^2+\gY^2}} = \frac{e}{\gL} \ ,
   \esp
 \ee
 where $\gY$ and $e$ denote the hypercharge and electromagnetic coupling constants, respectively.
 Finally $\zw$ depends on the VEVs and the other two mixing angles along with the gauge couplings, as
 \be\bsp
 &
  \tan(2\zw)=\frac{2 \gL \gR  \cos {\phw} \cos {\tw}(\cos^2 {\phw} k^2- \sin^2 {\phw}\vL^2)}
     {\gR^2\cos^2 {\tw} \vR^2-(\gL^2 - \gR^2\cos^2 {\phw} \cos^2 {\tw} ) \cos^2 {\phw} k^2 -
         (\gL^2 -  \gBL^2\cos^2 {\tw} \sin^2 {\phw}) \cos^2 {\phw} \vL^2}
           \ ,
\esp\label{eq:ewmix}\ee
As can be seen, $\zw$ is negligible for $k\ll v_R$.
Neglecting this $Z-Z^\prime$ mixing, the $Z$ and $Z^\prime$ boson
masses are given by
\be
  M_{Z} =  \frac{\gL}{2 \cos {\tw}} \ v
  \qquad\text{and}\qquad
  M_{Z^\prime} = \frac{1}{2\cos {\phw}} \sqrt{\gBL^2 \sin^2 {\phw} \cos^2 {\phw} \vL^2 +
     \gR^2 (\cos^4 {\phw} k^2 + \vR^2)} \ .
\label{eq:mz_mzp}\ee

\subsection{Scalar mass bounds }
\label{subsec:mass}

Before we embark on a detailed analysis of the $0\nu\beta\beta$ in this model,  we discuss the limits on the masses of the charged Higgs and gauge bosons in the model.
The lightest
charged Higgs boson in the ALRM is $H_2^\pm$, which is long-lived, so
that limits on its mass are
probed by searches for heavy stable charged particles. The
$H_2^\pm$ bosons are pair-produced via the Drell-Yan mechanism, in
proton-proton collisions at centre-of-mass energies of 7 TeV \cite{Aad:2011hz,Khachatryan:2011ts}, 8 TeV \cite{Chatrchyan:2013oca} and 13~TeV \cite{Aad:2020srt,Aaboud:2019trc,Aaboud:2017iio,Alimena:2019zri,Khachatryan:2016sfv,CMS:2016ybj} and
in electron-positron collisions at a centre-of-mass energy of 183~GeV \cite{Ackerstaff:1998si}.

The LEP results impose strict limits in the [45.9, 89.5]~GeV mass range~\cite{Ackerstaff:1998si}. The searches in 13~TeV LHC collisions exclude signal cross sections
ranging from 10 to 100~fb, the exact limit value depending on the
model, although direct
 limits are not straightforward to extract because of 
modelling of various detector effects which is complicated.
 Similar conclusions hold for 7 and 8~TeV LHC search
results~\cite{Chatrchyan:2013oca,Khachatryan:2011ts,Aad:2011hz}.

The heavier charged Higgs state $H_1^\pm$ is the one that would be relevant for the  $0\nu\beta\beta$ decay. This state would be constrained by more standard searches for additional Higgs states,
such as the one of ~\cite{Sirunyan:2019hkq}. Those searches are 
 targeting a specific production mode and a given decay channel.   For example,  Run~2 the analyses  in
 \cite{Sirunyan:2019hkq,Aaboud:2016dig}  investigated the
LHC sensitivity to  charged Higgs bosons decaying into $H^\pm\to\tau^\pm\nu_\tau$ \cite{Sirunyan:2019hkq}. There the results are interpreted within MSSM, and a lower limit $M_{H_1^\pm}>160$ GeV is quoted.
For heavier $H_1^\pm$ case,   the analyses of charged
Higgs boson production and decay in a $tb$ final state or heavy Higgs boson production in association with a $tb$ pair or a $W_Lb{\bar b}$ system have also been carried
out (see, {\it e.g.}, refs.~\cite{Sirunyan:2020hwv, Aaboud:2018cwk}). Limits on cross-sections times branching ratio are again obtained in the context of two benchmarks in the MSSM, again these are difficult to compare with our results since the couplings are completely different. 

Unlike in the ordinary LRSM, the charged right-handed gauge boson
$W_R$ couples to right-handed up-type quarks and charged leptons and their exotic
quarks and scotino partners, rather than the usual right-handed neutrinos and down quarks. In addition, there is no mixing between the SM $W_L$ boson and the its $SU(2)_{R^\prime}$ counterpart. Therefore, the limits on the $W_R$-boson mass, imposed from flavor violation,
 do not apply here, relaxing considerably constraints on its mass.  As the $W_R$ does not affect either $0\nu\beta\beta$ decay or leptogenesis, we do not discuss it further.

\section{Neutrinoless Double Beta Decay ($0\nu\beta\beta$) in ALRM}
\label{sec:discussion}

The half-life for various isotopes is related to the matrix elements and phase factor $G$. The matrix element has two parts, one corresponding to the nuclear currents ($\mathcal{M}$) and the other corresponding to the leptonic current  ($\eta$).  The expression for half-life is given by
\be
\frac{1}{T^{0\nu}_{1/2}}=\sum_i~G_i~\left|\eta_i{\cal M}_i\right|^2
\ee
where $i$ corresponds to standard as well as different BSM channels. 
Up to now, $0\nu\beta\beta$ decay has not been observed. The best half-life limits on $0\nu\beta\beta$ come from experiments on two isotopes, Ge$^{76}$ and Xe$^{136}$. The Heidelberg-Moscow collaboration gives 90\% C.L. limit of $ T^{0\nu}_{1/2}$($^{76}{\rm Ge}$) > $1.9 \times 10^{25}$  yrs \cite{KlapdorKleingrothaus:2000sn}, with the latest results on Germanium coming from GERDA  \cite{Agostini:2019hzm} and  MAJORANA \cite{Aalseth:2017btx} Collaborations quoting $9\times 10^{25}$ yrs and $2.7\times 10^{25}$ yrs, respectively.  Results from the Xenon experiments on the other hand are given by EXO-200 and KamLAND-ZEN quoting $T^{0\nu}_{1/2} (^{136}{\rm Xe})>1.8 \times 10^{25}  $ yrs \cite{Albert:2017owj} and $T^{0\nu}_{1/2}(^{136}{\rm Xe})>1.07 \times 10^{26}  $ yrs \cite{KamLAND-Zen:2016pfg}, respectively, both at 90\% C.L.

\begin{figure}[h]
	\begin{subfigure}{.23\textwidth}
\includegraphics[trim=2cm 21cm 15cm 20mm, clip,width=1\linewidth]{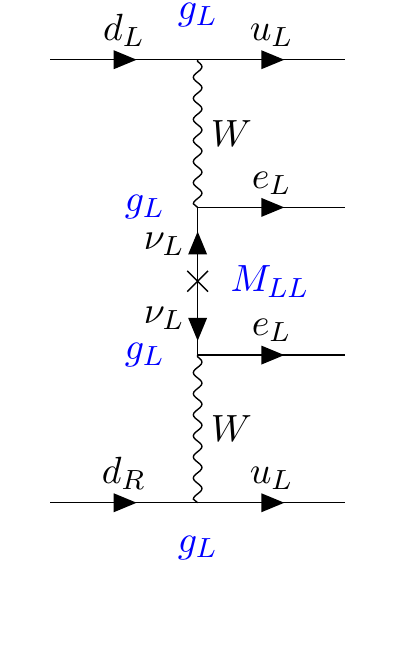}
        	\caption{ }
	\label{fig:feynmandiagWW}
	\end{subfigure}
	\begin{subfigure}{.23\textwidth}
     \includegraphics[trim=2cm 21cm 15cm 20mm, clip,width=1\linewidth]{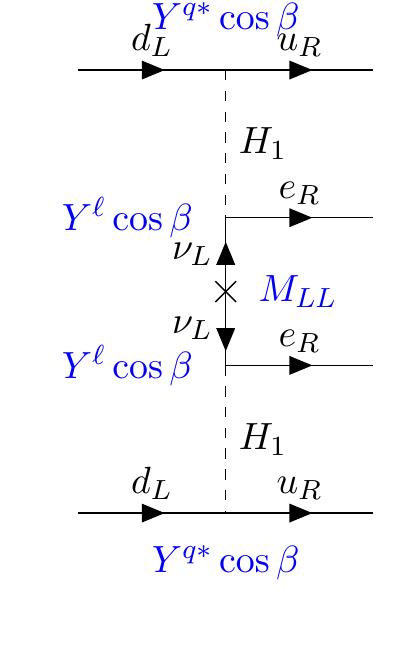}
        	\caption{ }
	\label{fig:feynmandiagHHeR1}
	\end{subfigure}
	\begin{subfigure}{.23\textwidth}
     \includegraphics[trim=2cm 21cm 15cm 20mm, clip,width=1\linewidth]{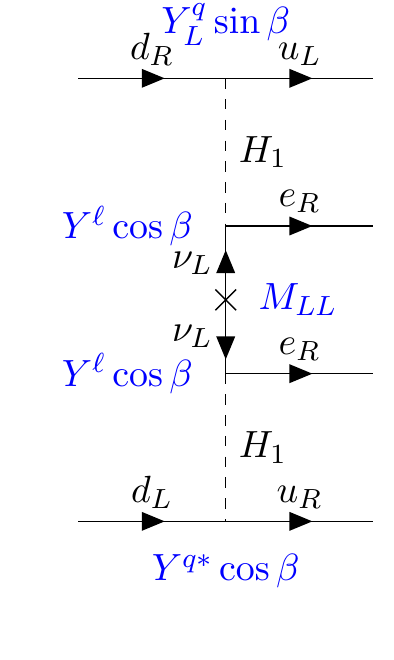}
        	\caption{ }
	\label{fig:feynmandiagHHeR2}
	\end{subfigure}
	\begin{subfigure}{.23\textwidth}
     \includegraphics[trim=2cm 21cm 15cm 20mm, clip,width=1\linewidth]{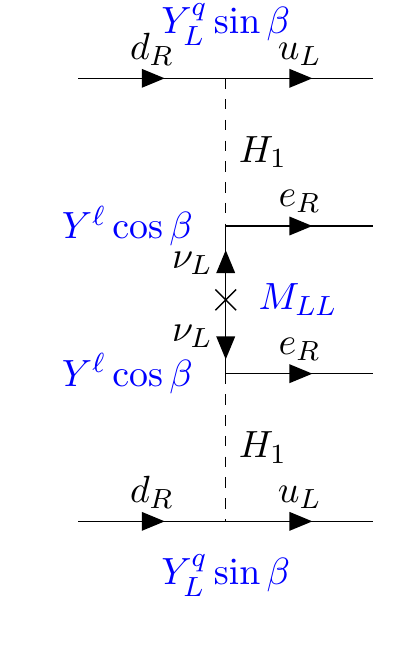}
        	\caption{ }
	\label{fig:feynmandiagHHeR3}
	\end{subfigure}	
\caption{\small Feynman diagrams contributing to the $0\nu\beta\beta$.  $W_L-W_L$ propagators with left-handed electrons,  and $H_1-H_1$ propagators with right-handed electrons. $M_{LL}=m_{\nu e}=\frac{(Y^e_Lv_L)^2}{m_N}$. Here and in the diagrams below, we depict the external particles and propagators in black, and couplings and mass insertions in blue.}  
	  \label{fig:feynmandiagWW_HH}
\end{figure}

\begin{figure}[h]
	\centering
	\begin{subfigure}{.32\textwidth}
     \includegraphics[trim=2cm 21cm 15cm 20mm, clip,width=0.7\linewidth]{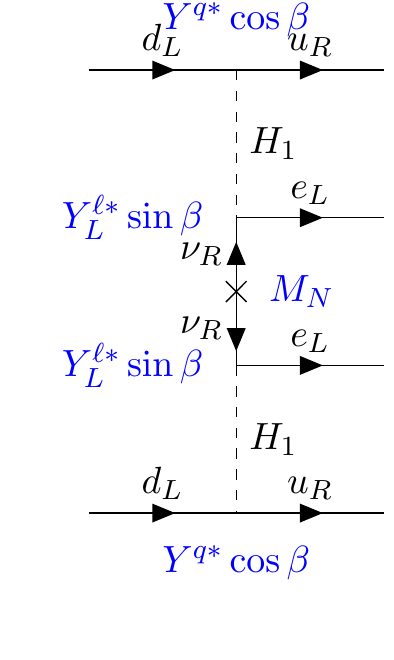}
        	\caption{ }
	\label{fig:feynmandiagHHeL1}
	\end{subfigure}
	\begin{subfigure}{.32\textwidth}
     \includegraphics[trim=2cm 21cm 15cm 20mm, clip,width=.7\linewidth]{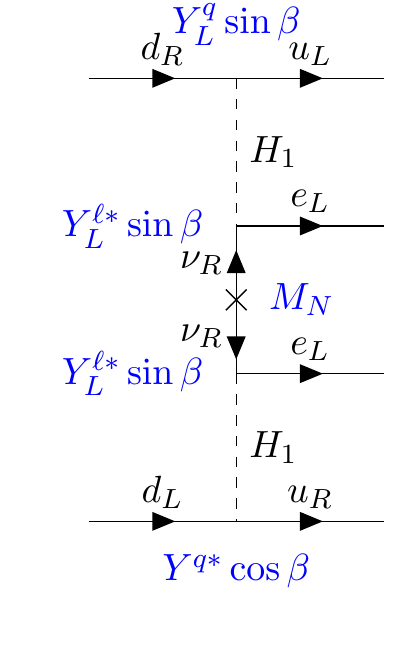}
        	\caption{ }
	\label{fig:feynmandiagHHeL2}
	\end{subfigure}
	\begin{subfigure}{.32\textwidth}
 \includegraphics[trim=2cm 21cm 15cm 20mm, clip,width=.7\linewidth]{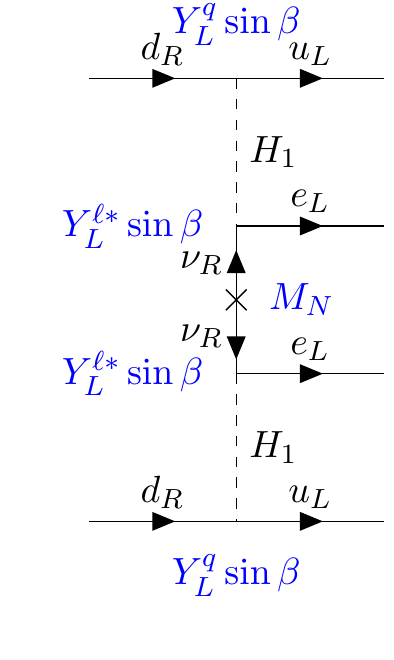}
        	\caption{ }
	\label{fig:feynmandiagHHeL3}
	\end{subfigure}	
\caption{\small Feynman diagrams contributing to the $0\nu\beta\beta$.  $H_1-H_1$ propagators and left-handed electron emission. The color code is as in \ref{fig:feynmandiagWW_HH}.}  
	  \label{fig:feynmandiagHHeL}
\end{figure}

\begin{figure}[h]
\centering
	\begin{subfigure}{.23\textwidth}
     \includegraphics[trim=2cm 21cm 15cm 20mm, clip,width=1\linewidth]{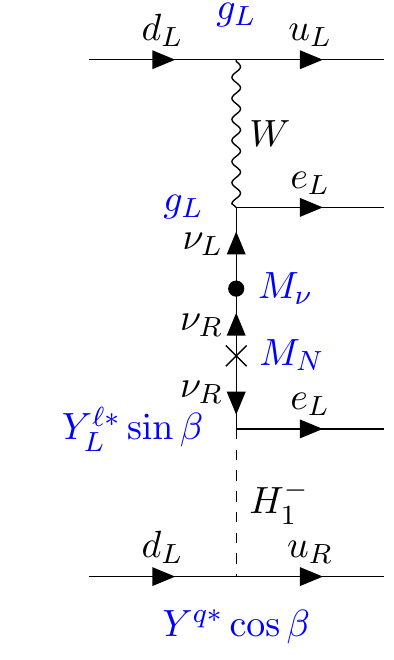}
        	\caption{ }
	\label{fig:feynmandiagWH1}
	\end{subfigure}
	\begin{subfigure}{.23\textwidth}
\includegraphics[trim=2cm 21cm 15cm 20mm, clip,width=1\linewidth]{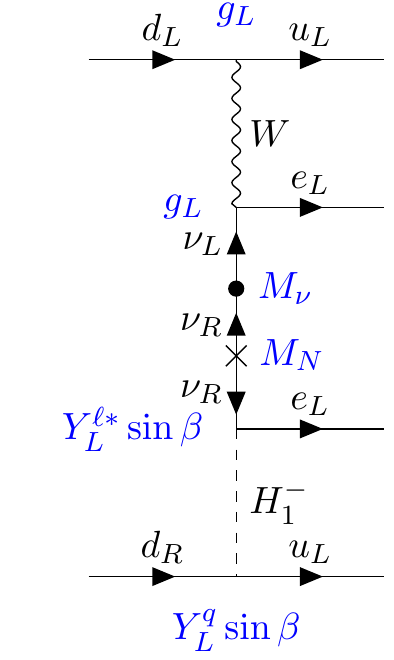}
        	\caption{ }
	\label{fig:feynmandiagWH2}
	\end{subfigure}
	\begin{subfigure}{.23\textwidth}
     \includegraphics[trim=2cm 21cm 15cm 20mm, clip,width=1\linewidth]{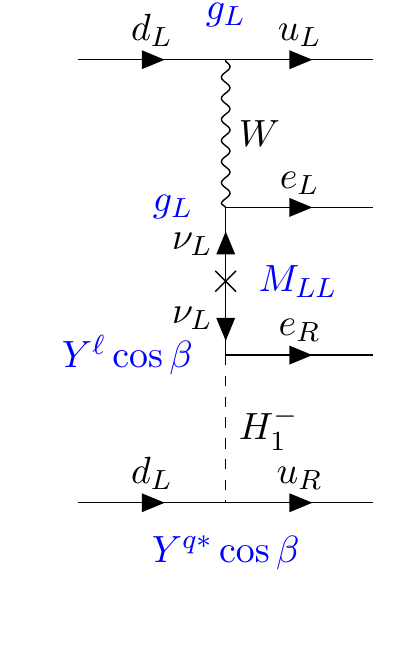}
        	\caption{ }
	\label{fig:feynmandiagWH3}
	\end{subfigure}
	\begin{subfigure}{.23\textwidth}
     \includegraphics[trim=2cm 21cm 15cm 20mm, clip,width=1\linewidth]{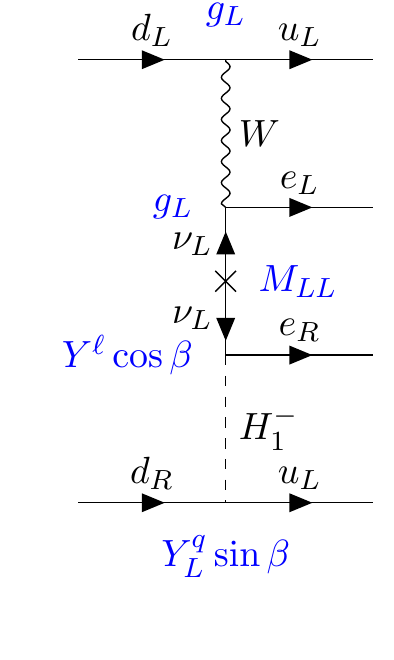}
        	\caption{ }
	\label{fig:feynmandiagWH4}
	\end{subfigure}
	\caption{\small Feynman diagrams contributing to the $0\nu\beta\beta$. $W_L-H_1$ propagators. The color code is as in \ref{fig:feynmandiagWW_HH}.}  
	  \label{fig:feynmandiagWH}
\end{figure}

Some of the relevant channels for $0\nu\beta\beta$ in ALRM are listed in figs.~\ref{fig:feynmandiagWW_HH} to \ref{fig:feynmandiagWH}. 
The usual  channel is given in \cref{fig:feynmandiagWW_HH} $(a)$ with the left-handed neutrinos and $W_L$ propagators. The other diagrams, \cref{fig:feynmandiagWW_HH} $(b),~(c),~(d)$ show the scalar-scalar mediated channel with two right-handed electrons in the final state, with the couplings explicitly indicated. Notice that the couplings are related to the mass of quarks and electrons, as given in 
\cref{eq:quarkmass} to \cref{eq:cosine}.  It is evident that the scalar mediated combinations are not significant. 
	
In \cref{fig:feynmandiagHHeL} channels with scalar-scalar propagator and two left-handed electrons in the final state are given. Again, we shall see that the contributions due to these diagrams are negligible.  On the other hand, the mixed scalar-vector propagator channels with Feynman diagrams in 
\cref{fig:feynmandiagWH} provide larger contributions, with \cref{fig:feynmandiagWH} $(d)$ contributing at par with the standard case for suitably chosen parameters.  

\subsection{Amplitudes and  parameters for various diagrams contributing to $0\nu\beta\beta$}
\label{subsec:appA}
Before writing down the required expressions for amplitudes and particle physics parameters for the diagrams, we define the transformation relation between flavor and mass eigenstates of neutrinos (both left- and right-handed) as,
\begin{eqnarray}
\nu_{L\alpha} &=& \mathcal{V}^{\nu\nu}_{\alpha i}\nu_i + \mathcal{V}^{\nu N}_{\alpha i}N_i \nonumber \\
\nu_{R\beta} &=& \mathcal{V}^{N\nu}_{\beta j}\nu_j + \mathcal{V}^{N N}_{\beta j}N_j \, ,
\label{eq:nmixing}
\end{eqnarray}
where $\alpha, \beta$ correspond to flavor indices $e, \mu, \tau$, while $i, j$ represent mass eigenstates 1, 2, 3. Also $\mathcal{V}$ denotes the rotation matrix between flavor and mass bases. The estimation for amplitudes as well as dimensionless particle physics parameters in context of various BSM scenarios can be found explicitly  in \cite{Mohapatra:1995yj,Babu:1995vh,Vergados:2012xy,Dev:2014xea}.  We divide the different channels into five categories, and we give the expressions for the amplitudes below.

\subsubsection{Vector-Vector ($W_L-W_L$) mediated diagrams  with $e_L-e_L$ emission}
\label{subsec:VVLL}
This so-called ``standard contribution'' in the literature appears in all  models allowing for $0\nu\beta\beta$, including that in the usual LRSM \cite{Barry:2013xxa} and is shown in the Feynman diagram in fig.\ref{fig:feynmandiagWW_HH}(a).   The amplitude for this process is given by
\begin{equation}
\mathcal{A}^{W_L W_L}_{LL, \nu_L} \sim G_F^2 \sum_i \left(\frac{\mathcal{V}^{{\nu\nu}^2}_{\alpha i}m_{\nu_i}}{p^2}-\frac{\mathcal{V}^{{\nu N}^2}_{\alpha i}}{m_{N_i}}\right)
\end{equation}
with the corresponding particle physics parameters 
\begin{equation}
\eta^{\nu, W_LW_L}_{LL, \nu_L}=  \frac{\sum_i\mathcal{V}^{{\nu\nu}^2}_{\alpha i}m_{\nu_i}}{m_e}
~~~~~~~
{\rm and}
~~~~~~~
\eta^{N_i, W_LW_L}_{LL, \nu_L}= -m_p \sum_i \frac{\mathcal{V}^{{\nu N}^2}_{\alpha i}}{m_{N_i}}.
\end{equation}

\subsubsection{Scalar-Scalar ($H_1-H_1$) mediated diagrams with $e_R-e_R$ emission}
\label{subsec:HHRR}
Among the two additional charged Higgs available in the ALRM framework, $H_2^-$ couples to the exotic down-type quark in the quark sector, and to the scotinos in the leptonic sector. $H_1^-$, on the other hand, could connect the quarks and the leptons in the standard spectrum. Possibilities of such helicity-flip scalar-scalar channels are identified from the early days as discussed in for example Refs. \cite{10.1143/PTPS.83.1,Mohapatra:1995yj}.
There are three different channels in the present case depending on the chirality of the quarks, as the Feynman diagrams in fig.~\ref{fig:feynmandiagWW_HH} $(b),~(c)$, $(d)$ show. The amplitudes and the corresponding particle physics parameters are given below,  
\begin{equation}
\mathcal{A}^{H_1 H_1}_{\sigma\sigma', \nu_L} \sim \frac{G_F^2}{g_L^4}~\frac{M_{W_L}^4}{M_{H_1}^4}~ \kappa_{ud}^2~ (Y^{l})^2 \cos^2 \beta\sum_i \left(\frac{\mathcal{V}^{{\nu\nu}^2}_{\alpha i}m_{\nu_i}}{p^2}-\frac{\mathcal{V}^{{\nu N}^2}_{\alpha i}}{m_{N_i}}\right)\, ,
\end{equation}
where $\sigma$ and $\sigma'$ denote the chirality of the outgoing $u$-quarks.  The dimensionless particle physics parameters are given by
\begin{equation}
\eta^{\nu_i, H_1H_1}_{\sigma\sigma', \nu_L}=\frac{M_{W_L}^4}{M^4_{H_1}}~\frac{\kappa_{ud}^2~ (Y^{l})^2 \cos^2 \beta}{g_L^4}~ \frac{\sum_i\mathcal{V}^{{\nu\nu}^2}_{\alpha i}m_{\nu_i}}{m_e},~~~~~
{\rm and}
\nonumber
\end{equation}
\begin{equation}
\eta^{N_i, H_1H_1}_{\sigma\sigma', \nu_L}=\frac{M^4_{W_L}}{M^4_{H_1}}~\frac{\kappa_{ud}^2~ (Y^{l})^2 \cos^2 \beta}{g_L^4} \sum_i \frac{-m_p~\mathcal{V}^{{\nu N}^2}_{\alpha i}}{m_{N_i}}\, .
\end{equation}
The coupling combination, $\kappa_{ud}^2$ depends on the chirality of the quarks. When both the $u$ quarks are right-handed  it is $(Y^{q\ast})^2\cos^2\beta$, and with both left-handed it is
$(Y^{q}_L)^2\sin^2\beta$, while for the mixed case of one left-handed and one right-handed quarks it is 
$Y^{q\ast}Y^q_L\sin\beta\cos\beta$.

\subsubsection{Scalar-Scalar ($H_1-H_1$) mediated diagrams  with $e_L-e_L$ emission}
\label{subsec:HHLL}
The amplitude of the scalar-scalar mediated process with right-handed neutrino exchange and the emission of two left-handed electrons, as shown in the Feynman diagrams in fig.~\ref{fig:feynmandiagHHeL}, is
\begin{equation}
\mathcal{A}^{H_1 H_1}_{\sigma\sigma', \nu_R} \sim \frac{G_F^2}{g_L^4}~\frac{M^4_{W_L}}{M^4_{H_1}}~ \kappa_{ud}^2~(Y^{l\ast}_L)^2\sin^2 \beta~ \sum_i \left(\frac{\mathcal{V}^{{N\nu}^2}_{\alpha i}m_{\nu_i}}{p^2}-\frac{\mathcal{V}^{{NN}^2}_{\alpha i}}{m_{N_i}}\right)\, .
\end{equation}
Similar to the previous case, the quark-chirality dependent coupling combination  $\kappa_{ud}^2$ is
$(Y^{q\ast})^2\cos^2\beta$ when both $u$-quarks are right-handed, $(Y^{q}_L)^2\sin^2\beta$ when both are left-handed, and $Y^{q\ast} Y^q_L\sin\beta\cos\beta$ in the mixed case. 
The dimensionless particle physics parameters for each of these channels (expressed in mass basis) can be written as
\begin{equation}
\eta^{\nu_i, H_1H_1}_{\sigma\sigma', \nu_R} =\frac{M^4_{W_L}}{M^4_{H_1}}~ \frac{\kappa_{ud}^2~(Y^{l\ast}_L)^2\sin^2 \beta}{g^4_L}~ \sum_i \frac{\mathcal{V}^{{N\nu}^2}_{\alpha i}m_{\nu_i}}{m_e}, ~~~~{\rm and}
\label{eq:whnuL1}
\end{equation} 
\begin{equation}
\eta^{N_i, H_1H_1}_{\sigma\sigma', \nu_R} = \frac{M^4_{W_L}}{M^4_{H_1}}~ \frac{\kappa_{ud}^2~(Y^{l\ast}_L)^2\sin^2 \beta}{g_L^4}~\sum_i \frac{-m_p~\mathcal{V}^{{NN}^2}_{\alpha i}}{m_{N_i}},
\label{eq:whnuL2}
\end{equation}
where $\eta^{X, H_1H_1}_{\sigma \sigma^\prime,\nu_R}$ denotes the contributions arising due to $X=\nu_i~ \text{or}~ N_i$ exchange in the mass basis.

\subsubsection{Vector-Scalar ($W_L-H_1$) mediated diagrams  with $e_L-e_L$ emission }
\label{subsec:WHLL}
Here, one of the quark currents connected with $W_L$ is left-handed, while the other current is helicity-flip interacting with the scalar. As discussed in Ref.~\cite{Mohapatra:1995yj, Babu:1995vh}, the amplitude is given by 
\begin{equation}
\mathcal{A}^{W_L H_1}_{\sigma\sigma', \lambda} \sim \frac{G_F^2}{g_L^2}~\frac{M^2_{W_L}}{M^2_{H_1}}~ \kappa_{ud}~Y^{l\ast}_L \sin\beta \sum_i \left(\frac{\mathcal{V}^{\nu\nu}_{\alpha i}\mathcal{V}^{N \nu \ast}_{\alpha i}}{\gamma . p} +   \frac{\mathcal{V}^{\nu N}_{\alpha i}\mathcal{V}^{N N \ast}_{\alpha i}}{m_{N_i}} \right),
\end{equation}
where the quark coupling, $\kappa_{ud}$ is $ Y^{q \ast} \cos\beta$ when one of the $u$-quarks is right-handed, while it is $Y^{q}_L  \sin\beta$ when both the $u$-quarks are left-handed. The dimensionless particle physics parameters are
\begin{equation}
 \eta^{\nu_i, W_L H_1}_{\sigma\sigma', \lambda}=  \frac{\kappa_{ud}~Y^{l\ast}_L \text{sin} \beta}{g_L^2}~\frac{M^2_{W_L}}{M^2_{H_1}} \sum_i \mathcal{V}^{\nu\nu}_{\alpha i}\mathcal{V}^{N \nu \ast}_{\alpha i},~~~~{\rm and}
\end{equation}
\begin{equation}
\eta^{N_i, W_L H_1}_{\sigma\sigma', \lambda} = \frac{\kappa_{ud}~Y^{l\ast}_L \text{sin} \beta}{g_L^2}~\frac{M^2_{W_L}}{M^2_{H_1}} \sum_i\mathcal{V}^{\nu N}_{\alpha i}\mathcal{V}^{N N \ast}_{\alpha i} \frac{\gamma.p}{m_{N_i}}.
\end{equation}

\subsubsection{Vector-Scalar ($W_L-H_1$) mediated diagrams  with $e_L-e_R$ emission :}
\label{subsec:WHLR}
The other two channels in this category are associated with the Feynman diagrams given in fig.\ref{fig:feynmandiagWH} $(c)$ and $(d)$. 
This gives the most promising contribution for the ALRM  considered here,  leading to helicity flipped lepton current. 
 The amplitudes are 
\begin{equation}
\mathcal{A}^{W_L H_1}_{\sigma\sigma', \nu_L} \sim \frac{G_F^2}{g_L^2}~\frac{M^2_{W_L}}{M_{H_1}^2}~\kappa_{ud}~Y^l \cos\beta \sum_i \left(\frac{\mathcal{V}^{{\nu\nu}^2}_{\alpha i}}{\gamma.p}-\mathcal{V}^{{\nu N}^2}_{\alpha i}\frac{\gamma.p}{m_{N_i}^2}\right)\, ,
\end{equation}
with $\kappa_{ud}= Y^{q \ast} \cos \beta$ when one of the $u$-quarks is right-handed, and $Y^{q }_L \sin \beta$ when both the $u$-quark are left-handed.  And the corresponding particle physics parameters are 
\begin{equation}
\eta^{\nu_i, W_LH_1}_{\sigma\sigma', \nu_L}=\frac{\kappa_{ud}~Y^l \cos\beta }{g_L^2}~\frac{M^2_{W_L}}{M^2_{H_1}}~ \sum_i \mathcal{V}^{{\nu\nu}^2}_{\alpha i}\frac{\gamma.p}{m_e}\, , ~~~~~{\rm and} 
\
\end{equation}
\begin{equation}
\eta^{N_i, W_LH_1}_{\sigma\sigma', \nu_L}=\frac{\kappa_{ud}~Y^l \cos\beta }{g_L^2}~\frac{M^2_{W_L}}{M^2_{H_1}}~\sum_i \mathcal{V}^{{\nu N}^2}_{\alpha i}\left(\frac{-m_p~\gamma.p}{m_{N_i}^2}\right)\,.
\end{equation}


\subsection{Numerical estimates}
\label{subsec:numerical}
The dimensionless particle parameters corresponding to different channels depend on the Yukawa couplings, the ratio the VEVs and the neutrino mixing matrix elements. The relevant combinations of Yukawa couplings and ratio of the VEVs, respecting the constraints discussed in \cref{sec:alrsm},  are given in \cref{table:Yukawa}, while all the mass parameters are given in \cref{table:mass}. 
\begin{table}[h]
\begin{subtable}{0.45\textwidth}
\begin{center}
\begin{tabular}{l|c}
\multicolumn{2}{c}{Yukawa Couplings}\\
\hline\hline
$Y^q\cos\beta$&$2.52\times 10^{-7}$\\ \hline
$Y^q_L\sin\beta$&$1.33\times 10^{-3}$\\ \hline
$Y^\ell\cos\beta$&$5.74\times 10^{-8}$\\ \hline
$Y^\ell_L\sin\beta$&$8.96\times 10^{-5}$\\
\hline\hline
\end{tabular}
\caption{ }
\label{table:Yukawa}
\end{center}
\end{subtable}
\begin{subtable}{0.45\textwidth}
\begin{center}
\begin{tabular}{l|c}
\multicolumn{2}{c}{Mass parameters}\\ \hline \hline
$m_p$&938 ~{\rm MeV}\\ \hline
$M_{W_L}$&80.4~{\rm GeV}\\ \hline
$M_{H_1}$&200 ~{\rm GeV}\\ \hline
$m_{N}$&10~{\rm TeV}\\ \hline
$m_\nu$&0.01~{\rm eV}\\ \hline
$\gamma\cdot p$&200~{\rm MeV}\\ \hline\hline
\end{tabular}
\caption{ }
\label{table:mass}
\end{center}
\end{subtable}
\caption{Combinations of Yukawa couplings and  ratio of the VEVs relevant to the particle physics parameters discussed in \cref{subsec:appA}.}
\label{table:couplings_masses}
\end{table}
The neutrino mixing matrix elements connect the mass eigenstates to the flavor eigenstates as in \cref{eq:nmixing}. In our case only the electron type neutrinos are relevant. Further, we have the unitarity condition
\begin{equation}
\sum_i{{\cal V}_{ei}^{\nu\nu}}^2 = \sum_i{{\cal V}_{ei}^{NN}}^2 = 1,
\end{equation}
and we considered 
\begin{equation}
\sum_i{\mathcal{V}_{ei}^{\nu\nu}}^2~m_{\nu_i} \sim 0.01~{\rm eV}\, , \qquad
\sum_i \frac{{\mathcal{V}^{NN}_{e i}}^2}{ m_{N_i}} \sim 0.1~{\rm (TeV)}^{-1}
\end{equation}
assuming close degeneracy of the neutrinos. This is easily followed by
$\mathcal{V}^{\nu N}_{e i} \sim \mathcal{V}^{N\nu}_{e i} \sim \left(\frac{m_\nu}{m_N}\right) \mathcal{V}^{\nu\nu}_{e i} \sim 10^{-15}$. We have, therefore, neglected the contributions proportional to $\mathcal{V}^{N\nu}_{e i} $, except when they are the leading contribution. In addition, we have considered the weak coupling constant, $g_L \sim 0.65$. With these values, the dimensionless particle physics parameters estimated for different channels are given in \cref{table:eta}. 
\begin{table}
\small
\begin{subtable}{0.33\textwidth}
\begin{center}
\begin{tabular}{r|c|c}
\multicolumn{3}{c}{\small $W_LW_L$ or $H_1H_1$ mediated with $\nu_L$ }\\ [1mm]\hline 
\cref{fig:feynmandiagWW}&$\left|\eta^{\nu_i, W_LW_L}_{LL, \nu_L}\right|$&$2 \times 10^{-8}$\\[1mm]\hline
\cref{fig:feynmandiagHHeR1}&$\left| \eta^{\nu_i, H_1 H_1}_{RR, \nu_L} \right|$&$ 1.3 \times 10^{-36}$\\[1mm] \hline
\cref{fig:feynmandiagHHeR2}&$\left|\eta^{\nu_i, H_1H_1}_{LR, \nu_L}\right| $&$6.6 \times 10^{-33}$\\ [1mm]\hline
\cref{fig:feynmandiagHHeR3}&$\left|\eta^{\nu_i, H_1H_1}_{LL, \nu_L}\right| $&$3.5 \times 10^{-29}$\\ [1mm] \hline \hline
\end{tabular}
\caption{ }
\label{table:eta1}
\end{center}
\end{subtable}
\begin{subtable}{0.33\textwidth}
\begin{center}
\begin{tabular}{r|c|c}
\multicolumn{3}{c}{\small $H_1H_1$ mediated with $\nu_R$  }\\ [1mm]\hline \hline
\cref{fig:feynmandiagHHeL1}&$\left|\eta^{N_i, H_1H_1}_{RR, \nu_R}\right|$&$1.4 \times 10^{-26}$\\[1mm] \cline{1-3}
\cref{fig:feynmandiagHHeL2}&$\left|\eta^{N_i, H_1H_1}_{LR, \nu_L}\right| $&$7.4 \times 10^{-23}$\\[1mm] \hline
\cref{fig:feynmandiagHHeL3}&$\left|\eta^{N_i, H_1H_1}_{LL, \nu_R}\right| $&$3.8 \times 10^{-19}$\\ [1mm]\hline\hline
\end{tabular}
\caption{ }
\label{table:eta2}
\end{center}
\end{subtable}
\begin{subtable}{0.33\textwidth}
\begin{center}
\begin{tabular}{r|c|c}
\multicolumn{2}{c}{\small $W_LH_1$ mediated }\\[1mm] \hline \hline
\cref{fig:feynmandiagWH1}&$\left| \eta^{\nu_i, W_L H_1}_{LR, \lambda} \right|$&$1.2 \times 10^{-25}$ \\ [1mm]\hline
\cref{fig:feynmandiagWH2}&$\left| \eta^{\nu_i, W_L H_1}_{LL, \lambda} \right|$&$ 6.4 \times 10^{-22}$  \\ [1mm]\hline
\cref{fig:feynmandiagWH3}&$\left|\eta^{\nu_i, W_LH_1}_{LR, \nu_L}\right| $&$3.2 \times 10^{-12}$ \\ [1mm]\hline
\cref{fig:feynmandiagWH4}&$\left|\eta^{\nu_i, W_LH_1}_{LL, \nu_L}\right|$&$1.7 \times 10^{-8}$ \\ [1mm]\hline \hline
\end{tabular}
\caption{ }
\label{table:eta3}
\end{center}
\end{subtable}
\caption{Estimated numerical values of dimensionless particle physics parameters for the couplings and masses given in \cref{table:couplings_masses}. }
\label{table:eta}
\end{table}
It is clear that the only significant channel is the one with $W_L-H_1$ mixed propagators. The corresponding particle physics parameters given in \cref{eq:whnuL1} can be re-expressed as
\begin{equation}
\eta^{\nu_i, H_1H_1}_{LL, \nu_R} =\frac{m_d~\gamma\cdot p}{2M_{H_1}^2 }= 1.7\times 10^{-8}~\left(\frac{\gamma\cdot p}{200~{\rm MeV}}\right)~\left(\frac{200~{\rm GeV}}{M_{H_1}}\right)^2.
\end{equation}

\subsection{Calculation of Half-life}
\label{subsec:halflife}
The best known mechanism leading to $0\nu \beta \beta$  decay is through the exchange of a Majorana
neutrino between the two decaying neutrons.   Applying the standard nuclear theory methods based on the non-relativistic
impulse approximation,  the general  $0\nu \beta \beta$ half-life formula in
 the $s$-wave approximation
 can be written in terms of the nuclear matrix elements (NME) and the particle physics parameters as \cite{Bonnet:2012kh}
\begin{eqnarray}
\frac{1}{T^{0\nu}_{1/2}}&=&G_{01}~\left| {\cal M}_{\nu_L}^W~\eta_{\nu_L}^W\right|^2 + G^R_{HH}~\left| {\cal M}_{\nu_L}^H~\eta_{\nu_L}^H \right|^2 + 
G^L_{HH}~\left| {\cal M}_{\nu_R}^H~\eta_{\nu_R}^H \right|^2  \nonumber \\
&&+G^{LL}_{WH}~\left| {\cal M}_{\lambda}^{WH}~\eta_{\lambda}^{WH} \right|^2 
+G^{LR}_{WH}~\left| {\cal M}_{\nu_L}^{WH}~\eta_{\nu_L}^{WH} \right|^2,
\end{eqnarray}
where 
\bea
\eta_{\nu_L}^W &=& \eta^{\nu_i, W_LW_L}_{LL, \nu_L},\nonumber \\
\eta_{\nu_L}^H&=&\eta^{\nu_i, H_1H_1}_{RR, \nu_L}+\eta^{\nu_i, H_1H_1}_{LR, \nu_L}+\eta^{\nu_i, H_1H_1}_{LL, \nu_L} \nonumber\\
\eta_{\nu_R}^H&=&\eta^{N_i, H_1H_1}_{RR, \nu_R}+\eta^{N_i, H_1H_1}_{LR, \nu_R}+\eta^{N_i, H_1H_1}_{LL, \nu_R} \nonumber \\
\eta_{\lambda}^{WH} &=&\eta^{\nu_i, W_LH_1}_{LR, \lambda}+\eta^{\nu_i, W_LH_1}_{LL, \lambda} \nonumber \\
\eta_{\nu_L}^{WH} &=&\eta^{\nu_i, W_LH_1}_{LR, \nu_L}+\eta^{\nu_i, W_LH_1}_{LL, \nu_L} \nonumber \\
\eea
The first term corresponds to the standard contribution with $W_L-W_L$ mediation (fig.\ref{fig:feynmandiagWW_HH}(a)), which is widely studied in the literature. 
The corresponding phase space factors and the nuclear matrix elements for two different isotopes Ge$^{76}$ and Xe$^{136}$ \cite{Kotila:2012zza,Pantis:1996py} are given in  \cref{tableNME}.
\begin{table}[h]
\centering
\begin{tabular}{c|c|c}
\hline\hline
Isotope & $G_{01} \; (\text{yrs}^{-1})$  &  $\mathcal{M}_{\nu_L}^W $  \\
\hline
$ \text{Ge}-76$ & $5.77\times10^{-15}$ & $2.58-6.64$   \\
$ \text{Xe}-136$ & $3.56 \times 10^{-14}$ & $1.57-3.85$  \\
\hline\hline
\end{tabular}
\caption{Standard $0\nu\beta\beta$ phase space factors \cite{Kotila:2012zza} 
and nuclear matrix elements for the different exchange processes \cite{Pantis:1996py} used in the 
analysis.}
\label{tableNME}
\end{table} 
 With the particle physics parameters given in \cref{table:eta1}, this leads to 
\bea
T_{\frac{1}{2},\,{\scriptscriptstyle  W_L}}^{0\nu}(^{76}{\rm Ge})&=& (9.82 - 65.09)\times 10^{27} \text{yrs}, \nonumber \\
T_{\frac{1}{2},\,{\scriptscriptstyle W_L}}^{0\nu}(^{136}{\rm Xe})&=& (4.74 - 28.49)\times 10^{27} \text{yrs},
\label{eq:thalf_std}
\eea
where the range  corresponds to the range in the NME in \cref{tableNME}.
To get an estimate of these factors for the other channels involving the scalar exchange, we turn to the discussion in Ref.~\cite{Pas:2000vn}.  
 The most general Lorentz invariant Lagrangian can be written in terms of the nuclear and lepton currents  as 
\begin{eqnarray}
\mathcal{L} &=& \frac{G_F^{2}}{2m_p} \left[\epsilon_1 JJj + \epsilon_2 J^{\mu \nu}J_{\mu \nu} j  + \epsilon_3 J^{\mu}J_{\mu}j  + \epsilon_4 J^{\mu}J_{\mu \nu} j^{\nu} + \epsilon_5 J^{\mu}J j_{\mu} \right. \\ \nonumber
&&\left. \phantom{\frac{G_F^{2}}{2m_p}}+ \epsilon_6 J^{\mu}J^{ \nu} j_{\mu \nu} + \epsilon_7 J J_{\mu \nu} j_{\mu \nu} + \epsilon_8 J_{\mu \alpha} J_{ \nu \alpha} j_{\nu}^{\mu}\right]\, ,
\end{eqnarray}
where $m_p$ is the mass of proton and the scalar, vector as well as tensor currents at the hadronic vertices are given by $J= \bar{u}(1\pm \gamma_5)d$, $J^\mu= \bar{u}\gamma^\mu (1\pm \gamma_5)d$, $J^{\mu \nu}= \bar{u}\frac{i}{2}[\gamma^\mu , \gamma^\nu](1\pm \gamma_5)d$,  respectively.  The corresponding leptonic currents can be expressed as $j= \bar{e}(1\pm\gamma_5)e^c$, $j^\mu= \bar{e}\gamma^\mu (1 \pm \gamma_5)e^c$, $j^{\mu \nu}= \bar{e}\frac{i}{2}[\gamma^\mu , \gamma^\nu](1\pm \gamma_5)e^c$, respectively. Also this $\epsilon$’s are same as our estimated $\eta$’s.
In the UV complete scenario that we presented here, there are no tensor operators. As seen from the discussion in \cref{subsec:numerical}, among the scalar and vector current operators, the only significant contribution is from the $W_L-H_1$ channel with the emission of two electrons with opposite chirality (\cref{fig:feynmandiagWH3} and \cref{fig:feynmandiagWH4}). The relevant term in the above Lagrangian corresponding to this  contribution is
\begin{equation}
\mathcal{L} \supset \frac{G_F^{2}}{2 m_p}~ (\epsilon_5 J^{\mu}J j_{\mu}),
\end{equation}
with the corresponding phase factor, as  in \cite{Pas:2000vn}, given by
\be
G_{WH}^{LR}(^{76}{\rm Ge}) = \frac{(m_e R)^2}{8} G_{09}(^{76}{\rm Ge})= 2.66\times 10^{-14}~\text{yrs}^{-1},~~~~~~G_{WH}^{LR}(^{136}{\rm Xe}) =1.29\times 10^{-14}~\text{yrs}^{-1}\, 
\ee
where we have used $G_{09}(^{76}{\rm Ge})=3.3\times 10^{-10}$ and $G_{09}(^{136}{\rm Xe})=1.6\times 10^{-9}$ as given in Refs.~\cite{Pas:2000vn, 10.1143/PTPS.83.1}. The nuclear radius, $R$ is taken as 10 fm, and $m_e$ is the mass of electron.
The nuclear matrix element for the two elements are, $\mathcal{M}_{\nu_L}^{WH} (^{76}{\rm Ge})=\mp 18.96$ (denoted as $\mathcal{M}_5$ in Ref.~\cite{Pas:2000vn}) and $\mathcal{M}_{\nu_L}^{WH} (^{136}{\rm Xe})=\mp 9.45$ \cite{Hirsch:1995ek}.   This leads to the half-life corresponding to the $W_L-H_1$ channel,
 \bea
 T_{\frac{1}{2},\,{\scriptscriptstyle WH}}^{0\nu}(^{76}{\rm Ge}) &=& 3.6 \times 10^{26}~\left(\frac{200~\text{MeV}}{\gamma \cdot p}\right)^2~\left(\frac{M_{H_1}}{200~\text{GeV}}\right)^4~ \text{yrs}, 
 \nonumber \\
T_{\frac{1}{2},\,{\scriptscriptstyle WH}}^{0\nu} (^{136}{\rm Xe})&=& 3.0 \times 10^{26}~\left(\frac{200~\text{MeV}}{\gamma \cdot p}\right)^2~\left(\frac{M_{H_1}}{200~\text{GeV}}\right)^4~ \text{yrs}\, .
\label{eq:thalf_WH}
\eea
These values are one to two orders of magnitude smaller than the standard channel given in \cref{eq:thalf_std}, while remaining safely within the current experimental limits quoted in the beginning of this section.

\section{Leptogenesis in ALRM}
\label{sec:leptogenesis}
Cosmological observations have definitively established  the preponderance of matter over antimatter. This asymmetry is measured relative to the number of photons $n_\gamma$ which can be extracted from observations and found to be \cite{Sakharov:1967dj,Dunkley:2008ie}
\begin{equation}
\Delta B\equiv \frac{n_B-n_{\bar B}}{n_\gamma} \sim 10^{-10} \, ,
\label{eq:DeltaB}
\end{equation}
 with $n_B,\, n_{\bar B}$ being the number of baryons and anti-baryons, respectively.  The underlying conditions for successful baryogenesis were formulated by Sakharov~\cite{Sakharov:1967dj} and allow for a wide variety of mechanisms, among which leptogenesis~\cite{Fukugita:1986hr} is of special interest because it establishes a connection between the Baryon Asymmetry of the Universe (BAU) and the generation of light active neutrino masses.

To explain BAU one  must go beyond the SM, either by introducing new sources of CP
violation and new kind of out-of-equilibrium situations (through  the  decay of some new heavy particles), 
or by modifying the electroweak phase transition itself. In leptogenesis 
a lepton asymmetry is generated before the electroweak phase
transition, which then is converted into BAU in the presence of sphaleron-induced anomalous $B + L$
violating processes.  These convert all primordial lepton asymmetry  into a baryon asymmetry. A
realization of leptogenesis through the decay of out-of-equilibrium  heavy neutrinos transforming as singlets under the SM gauge
group was proposed in Fukugita and Yanagida \cite{Fukugita:1986hr}. The additional CP violation is provided by the Yukawa  couplings through interference between tree and one-loop  decay diagrams. The departure from thermal  equilibrium occurs when the Yukawa interactions are sufficiently low. The lepton number violation in this scenario is generated from the Majorana masses of the heavy neutrinos,  giving rise to lepton number violating decays of the right handed neutrinos:
\begin{eqnarray}
\nu_{iR} &\to & l_{iL}+\Phi ^\dagger \nonumber \\
\nu_{iR} &\to & l^c_{iL}+\Phi \, .
\end{eqnarray}
 As usual,  CP violation comes from the interference of tree
level and one-loop (vertex and self-energy) diagrams. In ALRM,  the field entering the heavy neutrino decay is $H_1^\pm$\footnote{The charged components of the bidoublet field also contribute, but their contribution is suppressed by very small Yukawa couplings.}. The CP asymmetry
parameter corresponding to the vertex type CP violation is given
by
\begin{eqnarray}
\epsilon_v^{i} &=& \frac {\sum_\alpha \left[\Gamma ( \nu_{iR} \to  l_{\alpha L}+ H_1^+ )- \Gamma (\nu_{iR} \to l^c_{\alpha L}+H_1^- )\right]}
 {\sum_\alpha \left[\Gamma (\nu_{iR} \to  l_{\alpha L}+H_1^+ )+\Gamma (\nu_{iR} \to  l^c_{\alpha L}+H_1^- ) \right]}  \nonumber \\
&=& - \frac{1} {8 \pi}  \sum_{j=2,3}  \frac  { {\Im} \left[  \sum_\alpha (h^\star_{\alpha i} h_{\alpha j} )  \sum_\beta (h^\star_{\beta i} h_{\beta j} ) \right] } {\sum_\alpha | h_{\alpha i} |^2 }
f_v \left(  \frac{m_{N_j}^2}{m_{N_i}^2}  \right)\, ,
\end{eqnarray}
where $f_v(x)= \sqrt{x} \left[1-(1-x)\ln \left(\frac{1+x}{x}\right) \right]$.
In addition,  CP violation is generated by the interference of the tree level diagram with the one-loop
self-energy diagram CP violation, which resembles  the CP violation due to the box diagram in $K^0-{\bar K}^0$ mixing. If the heavy neutrinos
decay in equilibrium, the CP asymmetry arising from the self-energy diagram due to one of the heavy neutrinos may cancel
against the asymmetry from the decay of another, to preserve unitarity. However, in out-of-equilibrium decay of
heavy neutrinos, the number densities of the two heavy neutrinos differ during their decay, and  this cancellation is no longer exact.  The CP asymmetry  parameter coming from the interference of tree level and one-loop self-energy diagram  is given by:
\begin{eqnarray}
\epsilon_s^i &=& \frac {\sum_\alpha \left[\Gamma (\nu_{iR} \to  l_{\alpha L}+ H_1^+  -  \nu_{iR} \to l^c_{\alpha L}+H_1^- )\right] } {\sum_\alpha \left[\ \Gamma (\nu_{iR} \to  l_{\alpha L}+H_1^+ + \nu_{iR} \to  l^c_{\alpha L}+ H_1^+ )\right] } \nonumber \\
&=&\frac {1} {8 \pi}  \sum_{j=2,3}  \frac { {\Im}  \left[ \sum_\alpha (h^\star_{\alpha i} h_{\alpha j} ) \sum_\beta (h^\star_{\beta i} h_{\beta j} ) \right] }
{ \sum_\alpha | h_{\alpha i} |^2 }    
  f_s \left( \frac{m_{N_j}^2}{m_{N_i}^2}  \right)\, ,
\end{eqnarray}
with $f_s=\frac{\sqrt{x}}{1-x}$. In the ALRM scenario considered here, 
the coupling is $h^\star_{\alpha i}=  (Y_{ L }^{l_\alpha \star}) \sin \beta \,  {\cal V}^{NN\star}_{\alpha i}$ (where we have assumed diagonal Yukawa couplings).
These formulas are valid when the neutrino masses are strongly hierarchical ($m_{N_1} \ll m_{N_2},~m_{N_3}$) with $m_{N_{2,3}}-m_{N_1} \gg \frac 12 \Gamma_{N_1, N_{2,3}}$. In such case, the CP-violation arises from the decay of the lightest neutrino,
and we have  $\epsilon_v \sim \epsilon_s$.  With this, from the required out-of-equilibrium condition, the lower bound on the right-handed neutrino mass is $m_{N_1}>10^8 $ GeV \cite{Hati:2018tge}. However, in our scenario, as discussed in the previous sections, we require $m_N$ to be in the 1 - 10 TeV range, and thus, this situation does not apply to our case. It has been shown \cite{Pilaftsis:2003gt} that the mass limits on right-handed neutrinos can be significantly relaxed if two right-handed neutrinos are almost degenerate (resonant leptogenesis) and in this case the masses $m_{N_1} \sim m_{N_2} $, which can now be in the TeV range.  The contribution from the self-energy will now dominate ($\epsilon_s \gg \epsilon_v$) with
\begin{equation}
\epsilon_s^i =  \frac  { \Im  \left[ \sum_\alpha (h^\star_{\alpha i} h_{\alpha j} ) \sum_\beta (h^\star_{\beta i} h_{\beta j} )\right]}
{ \left(\sum_\alpha |h_{\alpha i} |^2\right)~\left(\sum_\beta |h_{\beta j} |^2\right) }~  \frac {   (m_{N_i}^2-m_{N_j}^2)m_{N_i} \Gamma_{N_j}}{   (m_{N_i}^2-m_{N_j}^2)^2  + m^2_{N_i} \Gamma^2_{N_j} }\, , 
\label{eq:resonant_epsilon}
\end{equation}
where we have 
\be
\Gamma_{N_i}=\frac{ (h^\dagger h)_{ii}} {8 \pi} m_{N_{i}}\, , 
\ee
with the condition,
\begin{equation}
m_{N_1}-m_{N_2} \simeq \frac 12 \Gamma_{N_1, N_2}\, .  
\end{equation}
For the actual values of the masses and widths, this leads to
\begin{equation}
\frac {   (m_{N_i}^2-m_{N_j}^2)m_{N_i} \Gamma_{N_j}}{   (m_{N_i}^2-m_{N_j}^2)^2  + m^2_{N_i} \Gamma^2_{N_j} }\simeq \frac{1}{2}\, .
\end{equation}
To achieve baryogenesis by leptogenesis, the baryon asymmetry is related to the CP-violating parameter $\epsilon$ through the relation
 $\Delta B \lesssim 10^{-4} \epsilon_s$, including the washout effects \cite{Buchmuller:2004nz}.  From \cref{eq:resonant_epsilon}, this leads to 
 \begin{equation}
 \frac  { \Im  \left[ \sum_\alpha (h^\star_{\alpha i} h_{\alpha j} ) \sum_\beta (h^\star_{\beta i} h_{\beta j} )\right]}
{ \left(\sum_\alpha |h_{\alpha i} |^2\right)~\left(\sum_\beta |h_{\beta j} |^2\right) }\simeq 10^{-7}\, ,
\end{equation}
to get the required baryon asymmetry as in \cref{eq:DeltaB}.
To investigate further, we need to understand the mixing matrix elements of ${\cal V}^{NN}$.
Recently the T2K experiment has given a 3$\sigma$ confidence interval for the $\delta_{CP}$ in the light neutrino sector, which is cyclic and repeats every 2$\pi$, as [-3.41, -0.03] for the so-called normal mass ordering and [-2.54, -0.32] for the inverted mass ordering \cite{Abe:2019vii}, but for heavy right-handed neutrinos, we do not have any information.

Before proceeding, we note that unlike in the case of LRSM, where right-handed neutrino masses are related to $W_R$ masses (as they are both proportional to $v_R$), and this affects the wash-out efficiency factor \cite{Dev:2014iva}, in ALRM the $W_R$ and $m_N$ masses are independent. The $W_R$ boson couples to charged leptons and  scotinos (the exotic neutrinos, part of the right-handed lepton doublet), while $N$ is a singlet, whose (Majorana) mass is a parameter in the Lagrangian. The decay of the right-handed neutrino and the evolution of its number density are not influenced by $W_R$, and thus safe from the possible washout present in the LRSM case.

To choose a simple example for our case, assume only two right-handed neutrinos $N_1$ and $N_2$, which are quasi-degenerate,  contributing maximally to leptogenesis. Then, from the self-energy contribution involving the intermediate $N_2$ neutrino, the CP-asymmetry in \cref{eq:resonant_epsilon} then gives
\be
| \epsilon^{\nu{N_1}}_s | \simeq \frac12 \frac {\left|{\Im} \left[( h^\dagger h)^2_{12} \right] \right|}{(h^\dagger h)_{11} (h^\dagger h)_{22} }\, .
\ee
For our model
\bea
\epsilon^{\nu{N_1}}_s &\simeq &\frac12 \frac {{\Im} \left[
 \sum_{\alpha} {\cal V}^{NN \star}_{\alpha 1} {\cal V}^{NN}_{\alpha 2}~  |Y_{ L }^{l_\alpha}|^2\right]^2 } {\sum_\alpha \left(|Y_{ L }^{l_\alpha}|^2~ | {\cal V}^{NN}_{\alpha 1}|^2\right)~
   \sum_\beta \left(|Y_{ L }^{l_\beta}|^2~  | {\cal V}^{NN}_{\beta 2}|^2\right)}
   \nonumber \\
  &=& \frac 12 \frac {{\Im} \left[
 {\cal V}^{NN \star}_{1 1}  {\cal V}^{NN}_{1 2} ~|Y_{ L }^{l_1 }|^2  + {\cal V}^{NN \star}_{2 1}  {\cal V}^{NN}_{2 2} ~|Y_{ L }^{l_2 }|^2 
 +{\cal V}^{NN \star}_{3 1}  {\cal V}^{NN}_{3 2} ~|Y_{ L }^{l_3 }|^2 \right]^2  } 
 {\left(|Y^{l_1}_L|^2|{\cal V}^{NN}_{11}|^2+|Y^{l_2}_L|^2|{\cal V}^{NN}_{21}|^2+|Y^{l_3}_L|^2|{\cal V}^{NN}_{31}|^2\right)~
 \left(|Y^{l_1}_L|^2|{\cal V}^{NN}_{12}|^2+|Y^{l_2}_L|^2|{\cal V}^{NN}_{22}|^2+|Y^{l_3}_L|^2|{\cal V}^{NN}_{32}|^2\right)}\, . \nonumber \\ 
 \eea
It would be reasonable to assume that all Yukawa couplings are real, and the phases emerging from the right-handed neutrino mixing matrix, for Majorana neutrinos, can be parametrized as
\be
{\cal V}^{NN}=  \left( \begin{array}{ccc}  C_{12}C_{13} & S_{12} C_{13} & S_{13}e^{-i \delta_N} \\
-S_{12}C_{23}-C_{12}S_{23}S_{13}e^{i \delta_N} & C_{12}C_{23}-S_{12}S_{23}S_{13}e^{i\delta_N} &S_{23}C_{13} \\
S_{12}S_{23}-C_{12}C_{23}S_{13}e^{i \delta_N} & -C_{12}S_{23}-S_{12}C_{23}S_{13}e^{i\delta_N} &C_{23}C_{13} 
\end{array} \right) {\rm Diag} (1, e^{i \alpha_M/2}, e^{i \beta_M/2} )\, ,
\ee
where $S_{ij}=\sin\theta^N_{ij}, \, C_{ij}=\cos \theta^N_{ij}, i,j=1,2,3$, $\delta_N$ is the Dirac phase, and $\alpha_M, \beta_M$ are Majorana phases in the right-handed neutrino mass matrix. Evaluating $\epsilon_s$ in terms of the matrix elements
\begin{eqnarray}
 \epsilon^{\nu{N1}}_s &\simeq& \frac{ S_{13}^2C_{23} \left ( S_{12}^2 S_{13}+C_{12}^2S_{23}\right ) \left [ C_{23}\left (S_{12}^2S_{13}-C_{12}^2S_{23} \right ) +S_{12}C_{12}\left( S_{23}-C_{23}^2S_{12} \right ) \right ] }
 { \left ( S_{12} S_{13}- C_{12}C_{23}S_{13} \right )^2 \left ( C_{12} S_{23}+S_{12} C_{23} S_{13} \right )^2}
 \sin \delta_N \, ,
 \label{eq:epsilons}
\end{eqnarray}
where we took into account the ordering of Yukawa couplings $Y_L^\tau \gg Y_L^\mu \gg Y_L^e$, and approximated $\cos \delta_N \simeq 1$.
Thus,  leptogenesis imposes limits on the phases of the mixing matrix for right-handed neutrinos. The expression in \cref{eq:epsilons} depends sensitively on the mixing angles in the matrix. Requiring  $|\epsilon^{\nu{N1}}_s| <10^{-6}$, the expression is in general complicated, and we cannot draw any definite conclusions. However, we can estimate the restriction on the Dirac phase in the heavy neutrino mass mixing under some simplifying conditions: 
\begin{itemize}
\item 
For $C_{ij}, S_{ij} ={\cal  O}(10^{-1})$, one would require that the Dirac CP violating phase in the right-handed neutrino masses be, 
$\sin \delta_N \simeq 10^{-5}$.
\item  If the matrix is mostly diagonal, $C_{ij} \sim {\cal O}(1) >> S_{kl} \sim {\cal O}(10^{-2})$, $\sin \delta_N \simeq 10^{-6}$, 
\item
 If the mixing is significant and the mass mixing matrix highly non-diagonal,  $S_{ij} \sim {\cal O}(1) >> C_{kl} \sim {\cal O}(10^{-2})$, $\sin \delta_N \simeq 10^{-6}$, 
 \item
Finally, if the third right-handed neutrino $N_3$ is significantly heavier than the first two and decouples $S_{13} \sim S_{23}= {\cal O}(10^{-2})$, and the first two right-handed neutrinos mix maximally, taking $C_{12}\sim S_{12} \sim 1/\sqrt{2}$, and $(C_{12}-S_{12})^2\sim {\cal O}(10^{-4})$, $\sin \delta_N \simeq 6 \times 10^{-12}$.
\end{itemize}
These results are promising, because they indicate that, for quite small Dirac phase in the right-handed neutrino mass matrix, the model generates sufficient leptogenesis to satisfy the BAU constraint. Conversely, leptogenesis limits the phase in the right-handed neutrino mass to be quite small. Since in the ALRM the light neutrino masses are generated  through flavor breaking, rather than by the seesaw mechanism, this has no direct implication for the Dirac phases in the PMNS matrix.
\section{Conclusions}
\label{sec:conclusions}
We have explored the ALRM, an alternate way of extending the SM gauge group with a relatively low energy $SU(2)_R$ but without any additional symmetry with the left handed sector, either in gauge couplings or in the matter content. As has been shown in earlier studies \cite{Babu:1987kp} this opens a way to achieve grand unification within $E_6$, while providing a rather unusual model for Dark Matter and an unusual phenomenology also for the vector bosons at the LHC \cite{Frank:2019nid}. Indeed, the scotino \cite{Ma:2010us}, the partner of the electron in the right-handed doublet in this model can be a viable dark matter candidate and has been studied in several variants of the model \cite{Khalil:2009nb,Khalil:2010yt,Farzan:2012sa,Ashry:2013loa}. As detailed in Sec. \ref{sec:alrsm}, the effective intermediate scale model is $SU(3)_c\times SU(2)_L\times SU(2)_{R^\prime} \times U(1)_{B-L} \times U(1)_S$, {\it i.e.},  the model needs to be enhanced by a global $U(1)_S$.

In an earlier attempt to understand the role of $SU(2)_R$ extension in $0\nu\beta\beta$  decay, a conventional left-right symmetric model  was employed \cite{Majumdar:2018eqz}, where right-handed charged currents mediated by $W_R$ and doubly charged scalars coupling to the leptons, permitting the gauge boson fusion channel, provided interesting new contributions. By contrast here those channels are absent, however the scalar mediated channels become more promising as the relevant  Yukawa interactions have new properties. Especially, we find that  the induced VEV of the left-handed doublet scalar  ($v_L$) generates masses for the down-type quarks as well as masses for the light neutrinos, and so it plays a crucial role here.
Our study establishes  that these new scalar mediated channels contribute significantly to the $0\nu\beta\beta$ in the case of $^{76}$Ge and $^{136}$Xe, the two nuclei experimentally explored. We have found that the contributions to the half-life can be one order larger than the standard $W_L$ mediated channels in both the cases,  with $T_{\frac{1}{2}}^{WH}\gtrsim 3\times 10^{26}~{\rm yrs}$ for  charged scalar mass of $M_{H_1^\pm}=200$ GeV, the only parameter that is sensitive to this computation.  This is well within the sensitivity expected by future experiments \cite{Armengaud:2019loe, Abgrall:2017syy}. 

In addition, the Majorana nature of the neutrinos has the potential to provide the required baryon asymmetry of the universe, through the mechanism of leptogenesis. However unlike the usual LRSM here the $W_R$ does not couple to the right handed neutrino, and plays no role in the wash out process of the generated lepton asymmetry. Studying the resonant leptogenesis, which allows the heavy neutrinos to be in the TeV range, 
the required CP violation can be easily obtained, even for a small Dirac phase in the right-handed neutrino mass mixing matrix. Thus  the ALRM emerging from $E_6$ GUT models provides interesting phenomenological consequences for the Majorana nature of neutrinos and lepton number violation, through enhanced $0\nu\beta\beta$ decay and leptogenesis.

\bigskip
\noindent
{\Large \bf Acknowledgement}\\[5mm]
\label{acknowledgement}
We gratefully acknowledge Shastri Institute for a Canada-India collaboration grant (SRG 2017-18). PP thanks the Department of Physics, Concordia University for its hospitality during this project. The work of MF has been partly supported by NSERC through grant number SAP105354. SS is thankful to UGC for fellowship grant to support her research work.
%
\bibliography{biblio}

\end{document}